\documentclass[prd,preprint,a4paper,superscriptaddress,nofootinbib,11pt]{revtex4}
\usepackage{amsmath,amsfonts,amssymb}
\usepackage{txfonts}
\usepackage[T1]{fontenc}

\newcommand{\ah}{\ensuremath{\hat{\alpha}}}
\newcommand{\an}{\ensuremath{\alpha}}
\newcommand{\ano}{\ensuremath{\alpha}_0}
\newcommand{\Ah}{\ensuremath{\hat{\mathcal A}}}

\newcommand{\As}{\ensuremath{{\mathcal A}_\star}}

\newcommand{\bh}{\ensuremath{\hat{\beta}}}
\newcommand{\bn}{\ensuremath{\beta}}
\newcommand{\bno}{\ensuremath{\beta}_0}

\newcommand{\eno}{\ensuremath{\epsilon_0}}

\newcommand{\Hn}{\ensuremath{\mathcal{H}}}

\begin{document}

\begin{center}
{\bf  \Large Twists, realizations and Hopf algebroid structure of $\kappa$-deformed phase space\\}
 
 \bigskip
\bigskip

Tajron Juri\'c  {\footnote{e-mail:e-mail: tjuric@irb.hr}} \\  
Rudjer Bo\v{s}kovi\'c Institute, Bijeni\v cka  c.54, HR-10002 Zagreb,
Croatia \\[3mm]

Stjepan Meljanac {\footnote{e-mail: meljanac@irb.hr}},
 \\  
Rudjer Bo\v{s}kovi\'c Institute, Bijeni\v cka  c.54, HR-10002 Zagreb,
Croatia \\[3mm] 
 
Rina  \v{S}trajn {\footnote{e-mail: r.strajn@jacobs-university.de}},
\\
Jacobs University Bremen, 28759 Bremen, Germany\\[3mm]

\end{center}
\setcounter{page}{1}


{ The quantum phase space described by Heisenberg algebra possesses undeformed Hopf algebroid structure. The $\kappa$-deformed phase space with noncommutative coordinates is realized in terms of undeformed quantum phase space. There are infinitely many such realizations related by similarity transformations. For a given realization we construct corresponding coproducts of commutative coordinates and momenta (bialgebroid structure). The $\kappa$-deformed phase space has twisted Hopf algebroid structure. General method for the  construction of twist operator (satisfying cocycle and normalization condition) corresponding to deformed coalgebra structure is presented. Specially, twist for natural realization (classical basis) of $\kappa$-Minkowski spacetime is presented. The cocycle condition, $\kappa$-Poincar\'{e} algebra and $R$-matrix are discussed. Twist operators in arbitrary realizations are constructed from the twist in the given realization using similarity transformations. Some examples are presented. The important physical applications of twists, realizations, $R$-matrix and Hopf algebroid structure are discussed.}

\bigskip
\textbf{Keywords:} noncommutative space, $\kappa$-Minkowski spacetime, twist deformation, $R$-matrix, realization, $\kappa$-deformed phase space, Hopf algebroid.\\
\textbf{PACS:} 02.40.Gh, 11.10Nx, 11.30.Cp, 02.20.Uw


\newpage

\section{Introduction}
The structure of spacetime at the Planck scale is unknown and represents an open problem which is under active research. Noncommutative (NC) spacetime emerged as a natural setting for capturing the essence of physical theories at very small distances.  In  \cite{Doplicher, Doplicher1} it was shown that the postulates of general relativity together with Heisenberg uncertainty principle  lead to spacetime uncertainty at the Planck scale $l_{\text{Planck}}$, i.e. $\Delta x_{\mu}\Delta x_{\nu}> l^2 _{\text{Planck}}$. The description of spacetime as a continuum of points (a smooth manifold) is an assumption no more justified at Planck scale. At this scale, it is then natural to relax the assumption of smooth spacetime and conceive spacetime as discretized manifold, most naturally described by noncommuative spacetime. This noncommutativity can be realized by promoting spacetime coordinates $x_{\mu}$ in to noncommuting  operators $\hat{x}_{\mu}$. The NC spacetime is also known to emerge as a low energy limit of certain quantum gravity models \cite{Doplicher, Doplicher1, kempf}. String theory \cite{Witten, boer} suggests that the spacetime at Planck length also leads to noncommutative spacetime.

In  NC spacetimes Lorentz symmetry is broken in the usual sense. Namely, Lorentz algebra remains undeformed,  its coalgebra changes , but in a way that we  still have the Hopf algebra of the starting symmetry group. A particularly interesting example of a Hopf algebra is $\kappa$-Poincar\'{e} algebra. $\kappa$-Poincar\'{e} algebra is describing the underlying symmetry of the effective NC quantum field theory that results from coupling quantum gravity to matter  fields after topological degrees of freedom of gravity are integrated out \cite{komentar,komentar1, komentar2}. It has been shown that the generic feature of  field theories on NC spaces (both for Moyal \cite{uv-ir,uv-ir1}, and $\kappa$-Minkowski \cite{uv-ir2}) is that interactions are highly non-local and non-linear leading to the so called UV/IR mixing, which is characterized by an interdependence between the high and low energy behavior \cite{douglas, Szabo}. 

We will deal with $\kappa$-Minkowski spacetime \cite{Lukierski-1}-\cite{MKJj}. $\kappa$-Minkowski spacetime is a Lie algebraic deformation of Minkowski spacetime, where $\kappa$ is the deformation parameter usually associated with quantum gravity scale. Investigations trying to obtain bound on deformation parameter, supporting this claim, were carried out in \cite{bounds, bgmp10, harisiva, bounds3, hajume}. The symmetries of $\kappa$-Minkowski spacetime are encoded in the $\kappa$-Poincar\'{e}-Hopf algebra. Generalized Poincar\'{e} algebras related to $\kappa$-Minkowski spacetime were considered in \cite{Kovacevic222}.
Constructions of physical theories on $\kappa$-Minkowski spacetime  lead to new interesting properties, such as,  modification of particle statistics \cite{kappaSt}-\cite{Gumesa}, deformed Maxwell's equations \cite{h,hjm11}, Aharonov-Bohm problem \cite{andrade} and quantum gravity effects \cite{dolan, solo, bgmp10, hajume, BTZ}.  Deformation of quantum mechanics and especially effects on hydrogen atom were also considered \cite{kuprijanov, kuprijanov1, harisiva}. The construction of QFT's on NC spaces is of immense importance and is still under investigation \cite{klm00}-\cite{mstw11}. $\kappa$-Minkowski spacetime is also related to doubly-special (DSR) and deformed relativity theories \cite{Amelino-Camelia-1, Kowalski-Glikman-1, Amelino-Camelia-2, Kowalski-Glikman-3, Kowalski-Glikman-2, bojowald}.

DSR theories are a set of models that provide a kinematical framework where Planck length is incorporated as a new fundamental invariant, along with the speed of light. The relativity postulates are minimally reformulated in order to allow this new invariant. In this fashion one avoids the necessity for singling out a preferred inertial frame, so that the concept of observer independence is retained, with $\kappa$-Minkowski spacetime providing the coordinate background for putting DSR to test. In due course some authors pointed towards certain inconsistencies and seeming paradoxes which DSR theories inevitably carry with them \cite{hosen}. The resolution of these problems was taken up in the recently proposed framework of relative locality \cite{rel-loc, rel-loc1, rel-loc2}. It relies on the concept of invariant phase space and idea that the momentum space might be curved. $\kappa$-Minkowski spacetime (invariant under $\kappa$-Poincar\'{e} algebra) was shown to emerge from the application of this idea, and can thus be used as a background on which one might develop implications of relative locality \cite{real-loc}.

It is known that the deformations of the symmetry group  can be realized through the application of the Drinfeld twist on that symmetry group \cite{Drinfeld, Drinfeld1, Majid-1}.  The main virtue of the twist formulation is that the deformed (twisted) symmetry algebra is the same as the original undeformed one and the only thing that changes is the coalgebra structure which then leads to the same free field structure as the corresponding commutative field theory \cite{chaichan}. The information about statistics is encoded in the $R$-matrix. In case of $\kappa$-Poincar\'{e} Hopf algebra, $R$-matrix can be expressed in terms of Poincar\'{e} generators only, which implies that the states of any number of identical particles can be defined in a $\kappa$-covariant way \cite{rmatrix,  kappaSt, kappaSt1, kappaSt2, kappaSt3, young}.

One of the ideas presented by the group of Wess et al. \cite{a4,a3} is that the symmetries of general relativity, i.e. the diffeomophisms, are considered as the fundamental objects and are deformed using twist \cite{Aschieri,a1}.  Given a twist $\mathcal{F}$ one can construct noncommutative star product. In this way, the algebra of noncommutative functions, tensor fields, exterior forms and diffeomorphisms is obtained. They also developed the notion of infinitesimal diffeomorphism and the corresponding notion of deformed Lie algebra. The generalization of the diffeormorphism symmetry is formulated in the language of Hopf algebras, a setting suitable for studying quantization of Lie groups and algebras. Physical applications of this approach are investigated in \cite{a2}, and especially for black holes in \cite{s2}. Our main motivation is to generalize the ideas of the group of Wess et al. to the notion of the Hopf algebroid \cite{Lu, Xu, Bem, mali} and to construct both QFT and gravity in Hopf algebroid setting, which is more general and it seems more natural since it deals with the  whole phase space \cite{rmatrix}.

There have been claims in the literature \cite{Borowiec-1} stating that $\kappa$-Poincar\'{e}-Hopf algebra could not be obtained from  cocycle twist, since $\kappa$-Poincar\'{e}-Hopf algebra is a quantum deformation of Drinfeld-Jimbo type corresponding to inhomogeneous $r$-matrix and that the universal $r$-matrix for $\kappa$-Poincar\'{e}-Hopf algebra is not known \cite{Lukierski-3, l1}. The Abelian twists \cite{Meljanac-4, ms06, Govindarajan-1}  and Jordanian twists \cite{Borowiec-3}  compatible with $\kappa$-Minkowski spacetime were constructed, but the problem with these twists is that they can not be expressed in terms of the Poincar\'{e} generators and the coalgebra runs out into $\mathcal{U(\mathfrak{igl}(\text{4}))}\otimes\mathcal{U(\mathfrak{igl}(\text{4}))}$.

 In a recent paper \cite{mali} we have demonstrated that the key for resolving these problems is to analyze the whole quantum phase space $\mathcal{H}$ and its Hopf algebroid structure.  We have used the Abelian twist, satisfying cocycle condition. This twist  is not an element of $\kappa$-Poincar\'{e}-Hopf algebra, but an element of  $\mathcal{H}\otimes\mathcal{H}$. By applying the twist to the Hopf algebroid structure of  quantum phase space $\mathcal{H}$ we obtained the Hopf algebroid structure of $\kappa$-deformed phase space $\hat{\mathcal{H}}$. Moreover,  this twist also provides the correct Hopf algebra structure of $\kappa$-Poincar\'{e} algebra when applied to the generators of rotation, boost and momenta. In \cite{mali} we have explicitly used the bicrossproduct basis (which corresponds to right ordering). In \cite{rmatrix}, authors have used deformations of the Heisenberg algebra (quantum phase space) and coalgebra by twist. They present a type of tensor exchange identities and show that the introduced coalgebra is compatible with them. Also, they give coproducts for the Poincar\'{e} generators, proposing two new methods of calculation. Finally, the exact form of the universal $R$-matrix for the deformed Heisenberg algebra and especially $\kappa$-Poincar\'{e} Hopf algebra is presented.

In this paper we present the construction of the twist for arbitrary realization and corresponding Hopf algebroid structure. A systematic, perturbative method for calculating twist in arbitrary realization of $\kappa$-deformed phase space is elaborated. 
We point out that for natural realization (also known as classical basis) there is no construction for the twist operator in the literature. Here we are presenting the expression for the twist in natural realization up to the third order in deformation parameter. We prove the cocycle condition and show that this twist leads to $\kappa$-Poincar\'{e}-Hopf algebra and we calculate the corresponding $R$-matrix.

Furthermore, we  show the relation between any two realizations via similarity transformation and give a method for calculating twist in a given realization, by using a twist in one particular realization and similarity transformation. Therefore, it is clear that if one knows the twist operator in one particular realization, using similarity transformation one can generate the twist operator in any realization. All such twists satisfy the cocycle and normalization conditions. Our general methods are demonstrated on specific examples, such as, left covariant, left noncovariant and natural realization.

In section II. we present and elaborate the structure of $\kappa$-deformed phase space $\hat{\mathcal{H}}$ and its realization through quantum phase space $\mathcal{H}$. We define the twist operator $\mathcal{F}$. Here we  develop the method for calculating twist in any realization of $\kappa$-deformed phase space and corresponding deformed coproduct (bialgebroid) structure. This twist satisfies cocycle condition. In section III. we explicitly apply the method developed in section II. for natural realization (up to the third order). Furthermore, using tensor relations, we prove that twist satisfies cocycle condition, calculate the $R$-matrix and show that this twist leads to $\kappa$-Poincar\'{e}-Hopf algebra. In section IV, we first give the structure of quantum phase space and explicitly show the relation between  generators, coproducts, star products and twists in arbitrary realizations (corresponding to different bases in quantum phase space) via similarity transformation. Then,  two realizations of $\kappa$-deformed phase space and  the formula that relates  twist operators in different realizations of $\kappa$-deformed phase space are given and discussed. In section V  the method developed in section IV. is illustrated in specific realizations. In section VI. discussion and possible physical applications of our approach are outlined. Finally, in the appendices, we give the definition and properties of undeformed Hopf algebroid, twisted Hopf algebroid and outline the Hopf algebroid structure of $\kappa$-deformed phase space $\hat{\mathcal{H}}$.

\section{Twist from realization} \label{2}
\subsection{$\kappa$-deformed phase space}
We start with $\kappa$-Minkowski spacetime defined by NC coordinates $\left\{\hat{x}_{\mu}\right\}$ ($\mu=0,1,2,3$) satisfying 
\begin{equation}\label{kappa}
[\hat{x}_{\mu},\hat{x}_{\nu}]\equiv \hat{x}_{\mu}\hat{x}_{\nu}-\hat{x}_{\nu}\hat{x}_{\mu}  =i(a_{\mu}\hat{x}_{\nu}-a_{\nu}\hat{x}_{\mu}),
\end{equation}
where $a_{\mu}=(a_{0},\vec{0})$.
Let us consider the realization of $\hat{x}_{\mu}$ in terms of commutative coordinates $x_{\mu}$ and momenta $p_{\mu}$ of the form
\begin{equation}\label{xrealization}
\hat{x}_{\mu}=x_{\alpha}\varphi^{\alpha}_{\   \mu}(p).
\end{equation}
Commutative coordinates $x_{\mu}$ and momenta $p_{\mu}$ generate the Heisenberg algebra $\mathcal{H}$ (i.e. quantum phase space) satisfying the following relations:
\begin{equation}\begin{split}\label{H}
&[x_{\mu},x_{\nu}]\equiv x_{\mu}x_{\nu}- x_{\nu}x_{\mu} =0,\\
&[p_{\mu},p_{\nu}]\equiv p_{\mu}p_{\nu}-p_{\nu}p_{\mu} =0 , \\
&[p_{\mu},x_{\nu}]\equiv p_{\mu}x_{\nu}-x_{\nu}p_{\mu} =-i\eta_{\mu\nu}\ 1,
\end{split}\end{equation}
where $\eta_{\mu\nu}=\text{diag}(-,+,+,+)$. 
The quantum phase space $\mathcal{H}$ is defined as an free unital algebra generated by $x_{\mu}$ and $p_{\mu}$, divided by the ideal generated by relations in eq. (\ref{H}). We choose bases element in $\mathcal{H}$ to be normally ordered monomials, i.e. coordinates $x_{\mu}$ are left from the momenta $p_{\mu}$, and write symbolically  $\mathcal{H}=\mathcal{A}\ \mathcal{T}$, where $\mathcal{A}$ is an unital commutative algebra generated by $x_{\mu}$ and $\mathcal{T}$ is an unital commutative algebra generated by $p_{\mu}$, $\mathcal{T}=\mathcal{C}[[p]]$. $\mathcal{H}$ is not a Hopf algebra\footnote{ In \cite{topan} Heisenberg algebra is defined by $[p_{i},x_{j}]=-i\hbar\delta_{ij}$, where $\hbar$ is treated as a central element in the Heisenberg algebra. Then, in the framework of ``hybrid quantization'', Heisenberg algebra could be treated as a Hopf algebra, but only in a weak sense. Namely, the generator $\hbar$ can be identified with the  identity operator only weakly $\hbar\approx1$. Obviously, when trying to encode physical system within Hopf algebra structure of Heisenberg algebra $\mathcal{H}$ one is led to inconsistencies. }, but it has Hopf algebroid structure over the base algebra $\mathcal{A}$ (see Appendix A).
Functions $\varphi^{\alpha}_{\   \mu}(p)$ in eq.(\ref{xrealization}) have to satisfy:
\begin{equation}\label{cond}
\frac{\partial \varphi^{\alpha}_{\   \mu}}{\partial p^{\beta}}\varphi^{\beta}_{\   \nu}-\frac{\partial \varphi^{\alpha}_{\   \nu}}{\partial p^{\beta}}\varphi^{\beta}_{\   \mu}=a_{\nu} \varphi^{\alpha}_{\   \mu}-a_{\mu} \varphi^{\alpha}_{\   \nu}.
\end{equation}
In the limit when $a_{\mu}\rightarrow0$ we have $\varphi^{\alpha}_{\   \mu}\rightarrow\delta^{\alpha}_{\ \mu}$. Eq. (\ref{cond}) possesses infinitely many solutions. For a given solution $\varphi^{\alpha}_{\   \mu}$ we can generate all other solutions by similarity transformations (which will be elaborated in section V).

 There exists an isomorphism between NC algebra $\cal \hat{A}$, generated by $\left\{\hat{x}_{\mu}\right\}$ and the algebra $\cal A_{\star}$, generated by commutative coordinates $\left\{x_{\mu}\right\}$ but with star multiplication $\star$. Star product  between two elements $f(x)$ and $g(x)$ of $\cal A_{\star}$ is defined as
\begin{equation}\label{star}
f(x)\star g(x)=\hat{f}(\hat{x})\hat{g}(\hat{x})\triangleright 1,
\end{equation}
where $\hat{f}(\hat{x})$ and $\hat{g}(\hat{x})$ are elements of $\cal \hat{A}$, and the action $\triangleright$ is defined by
\begin{equation}\label{djelovanje}
x_{\mu} \triangleright f(x)=x_{\mu}f(x),\quad p_{\mu}\triangleright f(x)=-i\frac{\partial f}{\partial x^{\mu}}.
\end{equation}
The star product defined in (\ref{star}) is associative for any choice of realization $\varphi^{\alpha}_{\   \mu}(p)$.
From  this isomorphism between $\cal \hat{A}$ and $\cal A_{\star}$ it follows that $\varphi^{\alpha}_{\   \mu}(p)$ is invertible (and vice versa), that is 
 $\varphi^{\alpha}_{\   \mu}(\varphi^{-1})^{\mu}_{\   \beta}=\delta^{\alpha}_{\  \beta}$.

$\kappa$-deformed phase space is generated by $\hat{x}_{\mu}$ and $p_{\mu}$ which we denote as deformed Heisenberg algebra $\hat{\mathcal{H}}$.
We have the action (for more details see \cite{kovacevic-meljanac}) 
$\blacktriangleright\  : \hat{\cal H}\otimes\hat{\mathcal{A}}\mapsto\hat{\cal A}$, where, symbolically,  $\hat{\cal H}=\hat{\mathcal{A}}\mathcal{T}$, $\hat{\cal A}$ is a subalgebra of $\hat{\cal H}$ generated by $\hat{x}_{\mu}$ and $\cal T$ is a subalgebra of $\hat{\cal H}$ generated by $p_{\mu}$: 
\begin{equation}\begin{split}\label{crnodjelovanje}
\hat{x}_{\mu} \blacktriangleright \hat{g}(\hat{x})&=\hat{x}_{\mu}\hat{g}(\hat{x}),\quad p_{\mu}\blacktriangleright 1=0,  \\
&p_{\mu}\blacktriangleright \hat{x}_{\nu}=-i\eta_{\mu\nu}.\\
\end{split}\end{equation}
From the Leibniz rule 
\begin{equation}
\hat{x}_{\mu}\blacktriangleright\hat{f}(\hat{x})\hat{g}(\hat{x})=\hat{x}_{\mu}\hat{f}(\hat{x})\hat{g}(\hat{x})
\end{equation}
we obtain the coproduct $\Delta\hat{x}_{\mu}$
\begin{equation}\label{delxkapa}
\Delta\hat{x}_{\mu}=\hat{x}_{\mu}\otimes 1.
\end{equation}
Note that $\hat{\mathcal{H}}$ has Hopf algebroid structure over the base algebra $\hat{\mathcal{A}}$ (see \cite{mali} and Appendix C).

For any choice of realization $\varphi^{\alpha}_{\   \mu}(p)$ we can construct  corresponding coproducts $\Delta$ for $x_{\mu}$ and  $p_{\mu}$ using Leibniz rules  for $p\triangleright (f\star g)$ and $x\triangleright (f\star g)$ obtained from  the property $h_{1}h_{2}\triangleright f(x)=h_{1}\triangleright(h_{2}\triangleright f(x))$, where $h_{1},\ h_{2}\in \mathcal{H}$ (see \cite{rmatrix}) . This construction leads to Hopf algebroid structure \cite{Lu,Xu,Bem}. The coproduct $\Delta x_{\mu}$ can also be obtained from
\begin{equation}\begin{split}
\Delta x_{\mu}&=\Delta (\hat{x}_{\alpha}(\varphi^{-1})^{\alpha}_{\ \mu})= \Delta (\hat{x}_{\alpha})\Delta(\varphi^{-1})^{\alpha}_{\ \mu}\\
&=(\hat{x}_{\alpha}\otimes 1)\Delta (\varphi^{-1})^{\alpha}_{\ \mu}=(x_{\nu} \varphi^{\nu}_{\ \alpha}\otimes 1)\Delta (\varphi^{-1})^{\alpha}_{\ \mu}.
\end{split}\end{equation}
The  coproduct for $p_{\mu}$ can be constructed (see \cite{ms11}, \cite{batisti} and \cite{kapa-snyder, kapa-snyder1}) using 
\begin{equation}
e^{ip\hat{x}}\triangleright e^{iqx}=e^{iP_{\mu}(p,q)x^\mu}
\end{equation}
 and calculating the star product between two plane waves 
 \begin{equation}
 e^{ipx}\star e^{iqx}=e^{i\mathcal{D}_{\mu}(p,q)x^\mu},
 \end{equation}
 where $\mathcal{D}_{\mu}(p,q)=P_{\mu}(K^{-1}(p),q)$ and $K(p)=P(p,0)$. The function $P_{\mu}$ is uniquely  determined with the choice of realization 
 $\varphi^{\alpha}_{\   \mu}$ via 
 \begin{equation}
  \frac{\text{d} P_{\mu}(\lambda p,q)}{\text{d}\lambda}=p^{\alpha}\varphi_{\mu\alpha}(P),
  \end{equation}
   where $\lambda$ is a parameter. The function $\mathcal{D}_{\mu}(p,q)$ determines the momentum  addition rule $\mathcal{D}(p,q)=p\oplus q$ in momentum space, and from it we can extract the coproduct for momentum $\mathcal{D}_{\mu}(p\otimes 1, 1\otimes p)=\Delta p_{\mu}$.     Coproduct is a unique mathematical object for fixed deformation parameters $a_{\mu}$. Coproducts in different  realizations are related by similarity transformations (see section V. for details). When $a_{\mu}\rightarrow 0$  coproduct $\Delta$ reduces to $\Delta_{0}$:
\begin{equation}\label{primitive}
\Delta_{0}x_{\mu}=x_{\mu}\otimes 1, \quad \Delta_{0}p_{\mu}=p_{\mu}\otimes 1 + 1\otimes p_{\mu},
\end{equation}
where we generate an equivalence class in $\mathcal{H}\otimes\mathcal{H}$ by  the ideal $\mathcal{I}_{0}=\mathcal{U}_{+}(\mathcal{R}_{0})(\mathcal{A}\otimes 1)\Delta_{0}\mathcal{T}$, where $\mathcal{U}_{+}(\mathcal{R}_{0})$ is a universal enveloping algebra generated by $(\mathcal{R}_{0})_{\mu}$ but without the unit element, $(\mathcal{R}_{0})_{\mu}\equiv x_{\mu}\otimes 1 - 1\otimes x_{\mu}$ (for more details see \cite{rmatrix}) and $\Delta_{0} \mathcal{T}=\mathcal{C}[[\Delta_{0}p]]\subset\mathcal{T}\otimes\mathcal{T}$ is the image of $\mathcal{T}$ in $\mathcal{T}\otimes\mathcal{T}$. Elements $\mathcal{R}_{0}$ satisfy following properties
\begin{equation}\begin{split}
[x_{\mu}\otimes 1,  &\mathcal{R}_{0}]=0, \quad [\Delta_{0}p_{\mu}, \mathcal{R}_{0}]=0,\\
&[(\mathcal{R}_{0})_{\mu},(\mathcal{R}_{0})_{\nu}]=0.
\end{split}\end{equation}
Since Heisenberg algebra $\mathcal{H}$ can be written as $\mathcal{H}=\mathcal{A}\ \mathcal{T}$ it can be shown that $\Delta_{0}\mathcal{H}=\mathcal{U}(\mathcal{R}_{0})(\mathcal{A}\otimes1)\Delta_{0}\mathcal{T}/\mathcal{I}_{0}=[(\mathcal{A}\otimes1)\Delta_{0}\mathcal{T}+\mathcal{I}_{0}]/\mathcal{I}_{0}$ is an algebra isomorphic to $\mathcal{H}$ and  that $\Delta_{0}\mathcal{A}=(\mathcal{A}\otimes 1 + \mathcal{I}_{0}) / \mathcal{I}_{0}=(\mathcal{A}\otimes\mathcal{A}+\mathcal{I}_{0}) / \mathcal{I}_{0}$ is an algebra isomorphic to $\mathcal{A}$.

We can also define the counit $\epsilon_{0}$ by $\epsilon_{0}(h)=h\triangleright 1$ for every $h\in \mathcal{H}$. Coproduct $\Delta_{0}$ and  counit $\epsilon_{0}$ together lead to bialgebroid structure of quantum phase space.
Let us mention that the Heisenberg algebra $\mathcal{H}$ has also Hopf algebroid structure, for more details see Appendix A.

\subsection{Twist operator}
The relation between deformed coproducts $\Delta$ and undeformed  $\Delta_{0}$, discussed in the previous subsection, defines the bidifferential twist operator $\mathcal{F}$  via
\begin{equation}\label{t}
\Delta h=\mathcal{F}\Delta_{0} h\mathcal{F}^{-1},
\end{equation}
 for every $h\in \mathcal{H}$. Hence,
\begin{equation}\label{cotw}
\Delta x_{\mu}={\cal F} \Delta_{0}x_{\mu} {\cal F}^{-1} , \quad \Delta p_{\mu}={\cal F} \Delta_{0}p_{\mu} {\cal F}^{-1}.
\end{equation}
The star product is given by
\begin{equation}
f(x)\star g(x)=m_{\star}(f(x)\otimes g(x))=m\left({\cal F}^{-1}\triangleright(f\otimes g)\right)
\end{equation}
where $m$ is the multiplication map defined by $m(h_{1}\otimes h_{2})=h_{1}h_{2}$, and $m_{\star}$ is the multiplication map defined by $m_{\star}(h_{1}\otimes h_{2})=h_{1}\star h_{2}$,  $\forall h_{1},h_{2}\in \mathcal{H}$.
The $\star$-product does not change if $\mathcal{F}^{-1}\rightarrow\mathcal{F}^{-1}+\mathcal{J}_{0}$, where $\mathcal{J}_{0}=\mathcal{U}_{+}(\mathcal{R}_{0})\mathcal{H}\otimes\mathcal{H}$ is the right ideal with the property $m(\mathcal{J}_{0}\triangleright(f\otimes g))=0$. Note that  $f(x)g(x)=m(f\otimes g)=m_{\star}(\mathcal{F}\triangleright(f\otimes g))$ does not  change if $\mathcal{F}\rightarrow\mathcal{F}+\mathcal{J}$, where $\mathcal{J}$ is also a right ideal defined by $\mathcal{J}=\mathcal{U}_{+}(\mathcal{R})\mathcal{H}\otimes\mathcal{H}=\mathcal{F}\mathcal{J}_{0}$, where  $\mathcal{R}_{\mu}=\mathcal{F}(\mathcal{R}_{0})\mathcal{F}^{-1}$. The property of right ideal $\mathcal{J}$ is $m_{\star}(\mathcal{J}\triangleright(f\otimes g))=0$. For the right ideals $\mathcal{J}_{0}$ and $\mathcal{J}$ we have
\begin{equation}\begin{split}
&\mathcal{J}_{0}(\mathcal{H}\otimes\mathcal{H})=\mathcal{J}_{0} ,\quad \mathcal{J}(\mathcal{H}\otimes\mathcal{H})=\mathcal{J}\\
&\ \ \ \ \ \ \ \ \ \ \mathcal{J}=\mathcal{F}\mathcal{J}_{0}\mathcal{F}^{-1}=\mathcal{F}\mathcal{J}_{0}\\
& \ \ \ \ \ \ \ \ \mathcal{J}\mathcal{J}_{0}\subset\mathcal{J} , \quad \mathcal{J}_{0}\mathcal{J}\subset \mathcal{J}_{0}
\end{split}\end{equation}
Twist must satisfy : 
\begin{enumerate}
\item cocycle condition:  
\begin{equation}\label{CO}
(\mathcal{F}\otimes 1)(\Delta_{0}\otimes 1)\mathcal{F}= (1\otimes \mathcal{F}) (1\otimes \Delta_0)\mathcal{F},
\end{equation}
\begin{equation}
(\mathcal{F}^{-1}\otimes 1)(\Delta\otimes 1)\mathcal{F}^{-1}= (1\otimes \mathcal{F}^{-1}) (1\otimes \Delta)\mathcal{F}^{-1},
\end{equation}
\item normalization condition:  
\begin{equation}\label{NO}
m(\epsilon_{0}\otimes1)\mathcal{F}=1=m(1\otimes\epsilon_{0})\mathcal{F},
\end{equation}
\item  In the limit when $a_{\mu}\rightarrow 0$ we have  ${\cal F}\rightarrow 1\otimes 1$,
\end{enumerate}
where $\epsilon_{0}$ is the counit.

It can be shown that $\Delta \mathcal{H}=\mathcal{U}(\mathcal{R})(\mathcal{A}\otimes1)_{\mathcal{F}}\Delta\mathcal{T}/\mathcal{I}$ is an algebra isomorphic to $\mathcal{H}$ where $(\mathcal{A}\otimes 1)_{\mathcal{F}}=\mathcal{F}(\mathcal{A}\otimes 1)\mathcal{F}^{-1}$ and  $\mathcal{I}=\mathcal{F}\left(\mathcal{I}_{0}\right)\mathcal{F}^{-1}=\mathcal{U}_{+}(\mathcal{R})(\mathcal{A}\otimes 1)_{\mathcal{F}}\Delta\mathcal{T}$. It can be also shown that 
$\Delta \mathcal{A}=\mathcal{F}\left( \Delta_{0}\mathcal{A}\right) \mathcal{F}^{-1}=((\mathcal{A}\otimes 1)_{\mathcal{F}}+\mathcal{I})/\mathcal{I}$ is an algebra isomorphic to $\mathcal{A}$. Elements $\mathcal{R}_{\mu}$ satisfy the following properties 
\begin{equation}\begin{split}
[\Delta x_{\mu}, &\mathcal{R}]=0, \quad [\Delta p_{\mu}, \mathcal{R}]=0,\\
&[\mathcal{R}_{\mu},\mathcal{R}_{\nu}]=0.
\end{split}\end{equation}
For the ideals $\mathcal{I}_{0}$ and $\mathcal{I}$ we have 
\begin{equation}\begin{split}
&\mathcal{I}_{0}\Delta_{0}\mathcal{H}=\Delta_{0}\mathcal{H}\mathcal{I}_{0}=\mathcal{I}_{0}, \quad  \mathcal{I}\Delta\mathcal{H}=\Delta\mathcal{H}\mathcal{I}=\mathcal{I},\\
&\ \ \ \ \ \ \ \ \ \mathcal{I}\mathcal{I}_{0}\subset\mathcal{J} , \quad \mathcal{I}_{0}\mathcal{I}\subset\mathcal{J}_{0}
\end{split}\end{equation}

The twist operator $\mathcal{F}$ defined in (\ref{t}) is a mapping $\ \mathcal{F}:\Delta_{0} \mathcal{H}\mapsto \Delta \mathcal{H}$ and  $\mathcal{F}\in (\mathcal{H}\otimes \mathcal{H})/\mathcal{J}$, while its inverse $\mathcal{F}^{-1}$ is a mapping $\mathcal{F}^{-1}:\Delta \mathcal{H}\mapsto\Delta_{0} \mathcal{H}$ and  $\mathcal{F}^{-1}\in(\mathcal{H}\otimes \mathcal{H})/\mathcal{J}_{0}$. Hence,  $\Delta_{0}h:\Delta_{0}\mathcal{H}\mapsto\Delta_{0} \mathcal{H}$ and $\Delta h:\Delta\mathcal{H}\mapsto\Delta \mathcal{H}$. The twisted structure of quantum phase space $\mathcal{H}$ is also a Hopf algebroid over the base algebra $\hat{\mathcal{A}}$, where the elements of $\hat{\mathcal{A}}$ are taken in particular realization (see Appendix B).

For a given realization $\varphi^{\alpha}_{\   \mu}(p)$, eq.(\ref{xrealization}), and the corresponding coproducts $\Delta x$ and $\Delta p$ we  construct the twist $\cal F$ using perturbative methods. We  express $\Delta x$ and $\Delta p$ as  power series in the deformation parameter $a_{0}$
\begin{equation}\label{D}
\Delta x= \sum^{\infty}_{k=0}\Delta_{k}x, \quad \Delta p= \sum^{\infty}_{k=0}\Delta_{k}p,
\end{equation}
where $\Delta_{0}x$ and $\Delta_{0}p$ are given in (\ref{primitive}) and $\Delta_{k}x,\Delta_{k}p \propto a^{k}$. For the twist we have
\begin{equation}\label{f}
{\cal F} =\text{e}^{f}, \quad f=\sum^{\infty}_{k=1} f_{k},
\end{equation}
where $f_{k}\propto a^{k}x p^{k+1}$, symbolically. Using (\ref{cotw}) and comparing (\ref{D}) and (\ref{f}) order by order we have
\begin{equation}\begin{split}\label{formule}
&\Delta_{1}x=[f_{1},\Delta_{0}x]\\
&\Delta_{2}x=[f_{2},\Delta_{0}x]+\frac{1}{2}[f_{1},[f_{1},\Delta_{0}x]]\\
&\Delta_{3}x=[f_{3},\Delta_{0}x]+\frac{1}{2}\left([f_{1},[f_{2},\Delta_{0}x]]+[f_{2},[f_{1},\Delta_{0}x]]\right)+\frac{1}{3!}[f_{1},[f_{1},[f_{1},\Delta_{0}x]]]\\
&...\\
&\Delta_{k}x=[f_{k},\Delta_{0}x]+...+\frac{1}{k!}[f_{1},[f_{1},...[f_{1},\Delta_{0}x]]]\\
&...\\
\end{split}\end{equation}
and analogously for $\Delta p_{\mu}$. Note that for $\Delta_{1}x$ and $\Delta_{1}p$ we have
\begin{equation}\begin{split}\label{del1}
&\Delta_{1}x_{\mu}=a_{0}x_{\alpha}\left[ 1\otimes\frac{\partial [\varphi^{-1}]^{\alpha}_{\ \mu}}{\partial a_{0}}+\frac{\partial \varphi^{\alpha}_{\ \mu}}{\partial a_{0}}\otimes 1\right]_{a_{0}=0}\\
&\Delta_{1}p_{\mu}=a_{0}p_{\alpha}\otimes \left[\frac{\partial \varphi^{\ \alpha}_{\mu}}{\partial a_{0}}\right]_{a_{0}=0}
\end{split}\end{equation}

The important result is that associative star product implies the cocycle condition for twist $\mathcal{F}$ up to right ideal $\mathcal{J}$ (the complete proof will be presented elsewhere). This is the reverse statement to the one found in \cite{a3}, namely that the cocycle condition leads to associative star product. 

In \cite{Meljanac-4, ms06} is given the formula for operator $\mathcal{F}^{-1}$, in terms of $\Delta p_{\mu}$, that gives proper associative star product
\begin{equation}\label{:}
\mathcal{F}^{-1}=:\text{exp}\left(ix_{\alpha}(\Delta-\Delta_{0})p^{\alpha}\right):\ \ \ \in (\mathcal{H} \otimes \mathcal{H})/\mathcal{J}_{0},
\end{equation}
where $:\ \ :$ denotes normal ordering i.e. that $x\ \text{'s}$ are left from $p\ \text{'s}$. Using relations $(\mathcal{R}_{0})_{\mu}=x_{\mu}\otimes1-1\otimes x_{\mu}\equiv 0$ one can show that $\mathcal{F}^{-1}$ satisfies all the  requirements of the twist operator. This was demonstrated in \cite{Govindarajan-1} for an infinite class of Abelian twists. Note that using (\ref{del1}) and (\ref{:}) we can obtain $\mathcal{F}^{-1}$ up to the first order, in arbitrary realization 
\begin{equation}
\mathcal{F}^{-1}=1\otimes1+ia_{0}x^{\alpha}p_{\beta}\otimes \left[\frac{\partial \varphi^{\ \beta}_{\alpha}}{\partial a_{0}}\right]_{a_{0}=0}+O(a^{2}_{0}).
\end{equation} 

For different coproducts $\Delta x$ within the given equivalence class generated by $\mathcal{R}$ we can construct families of different twists, but they are all equivalent after using tensor relations $\mathcal{R}$. These tensor relations have to be used in the expansion of twist order by order. Hence, our method for calculating twist $\mathcal{F}$ from a given realization $\varphi^{\alpha}_{\ \beta}$ is well defined.

\section{Twist from natural realization of $\kappa$-Minkowski spacetime } \label{3}

An important example of a realization of the $\kappa$-Minkowski spacetime is the natural realization\cite{Meljanac-3, ms06, ms11}(or classical basis \cite{Kowalski-Glikman-1,natbor}). Important feature of natural realization is that momentum transforms vector-like under the undeformed Lorentz algebra.  In this  section we will demonstrate the method developed in previous section in order to construct the twist operator for natural realization. So far, the twist corresponding to natural realization is not known in the literature. For this case we use capital letters $X_\mu$ and $P_\mu$ to denote the commutative coordinates and the corresponding momentum operators. In terms of these, the noncommutative coordinates are given by
\begin{equation}
\hat{x}_\mu\equiv X_{\alpha}\Phi^{\alpha}_{\ \mu}(P)= X_\mu Z^{-1}-(a\cdot X) P_\mu, \label{hxnr}
\end{equation}
where $Z$ is the shift operator defined by $[Z,\hat{x}_\mu]=ia_\mu Z$ and, for the natural realization, given by\footnote{where we use the notation $A\cdot B\equiv A_{\mu}B^{\mu}$ and $A^2\equiv A_{\mu}A^{\mu}$ ($A$ and $B$ are arbitrary 4-vectors)} $Z^{-1}=(a\cdot P)+\sqrt{1+a^2P^2}$. The importance of this realization is given by the fact that generators of the undeformed Lorentz algebra in this case can be realized in the usual way and that $M_{\mu\nu}$ and $\hat{x}_{\mu}$ generate a Lie algebra:
\begin{equation}\label{MxLie}
[M_{\mu\nu},\hat{x}_{\lambda}]=-i(\eta_{\nu\lambda}\hat{x}_{\mu}-\eta_{\mu\lambda}\hat{x}_{\nu}+a_{\mu}M_{\nu\lambda}-a_{\nu}M_{\mu\lambda})
\end{equation}
also
\begin{equation}\begin{split}
[M_{\mu\nu},M_{\lambda\rho}]=-i(\eta_{\nu\lambda}M_{\mu\rho}&-\eta_{\mu\lambda}M_{\nu\rho}-\eta_{\nu\rho}M_{\mu\lambda}+\eta_{\mu\rho}M_{\nu\lambda})\\
[M_{\mu\nu},P_{\lambda}]&=\eta_{\nu\lambda}P_{\mu}-\eta_{\mu\lambda}P_{\nu}
\end{split}\end{equation}
In the following we demonstrate the procedure outlined in the previous section  for this case, i.e., for
\begin{equation}
\Phi_{\alpha\mu}(P)=\eta_{\alpha\mu}Z^{-1}-a_{\alpha}P_{\mu}.
\end{equation}
It can be checked that the inverse of this matrix is given by
\begin{equation}
\Phi_{\beta\gamma}^{-1}(P)= \eta_{\beta\gamma}Z +\frac{a_{\beta}P_\gamma Z}{\sqrt{1+a^2P^2}}.
\end{equation}
The coproduct of $P_\mu$ is known to be \cite{Kowalski-Glikman-1, Meljanac-3, ms06, kovacevic-meljanac}
\begin{equation}
\Delta P_\mu=P_\mu\otimes Z^{-1} +1\otimes P_\mu +ia_\mu P^{\alpha}Z\otimes P_{\alpha} -\frac{ia_\mu}{2}\square Z\otimes (a\cdot P),
\end{equation}
where $\square$ is the Casimir operator defined by $[M_{\mu\nu},\square]=0, [P_\mu, \square]=0$ and $[\square, \hat{x}_\mu]=-i2P_\mu$, and, for the natural realization, given by $\square=\frac{2}{a^2}\left(1-\sqrt{1+a^2P^2} \right)$. These expressions are inserted into
\begin{equation} \label{copXform}
\Delta X_\mu  =\left(X_{\alpha}\Phi^{\alpha\beta}\otimes 1\right) \left(\Delta \Phi_{\beta\mu}^{-1}\right),
\end{equation}
and expanded in the deformation parameter\footnote{we took that $a_{\mu}=(a_{0},\vec{0})$} $a_0$. The coproduct $\Delta \Phi_{\beta\mu}^{-1}$ is calculated using the homomorphism property of the coproduct. In the first order we get
\begin{equation}
\Delta X_\mu =X_\mu \otimes 1-X_\mu \otimes (a\cdot P)+(a\cdot X)\otimes P_\mu +O(a_0^2).
\end{equation}
The coproduct $\Delta P_\mu$ is also expanded to the first order in $a_0$
\begin{equation}
\Delta P_\mu =P_\mu \otimes 1 +1\otimes P_\mu +P_\mu \otimes (a\cdot P) -a_\mu P^{\alpha}\otimes P_\alpha +O(a_0^2),
\end{equation}
and we find $f_1$ by writing down the ansatz
\begin{equation}
f_1=-\alpha_1 (X\cdot P)\otimes (a\cdot P) -\alpha_2 (a\cdot X) P_\alpha \otimes P^{\alpha},
\end{equation}
where $\alpha_{1}$ and $\alpha_{2}$ are dimensionless parameters. Then we  require (see eq.(\ref{formule})) 
\begin{equation}\begin{split}
[f_1, X\otimes 1]&= -X_\mu \otimes (a\cdot P) +(a\cdot X) \otimes P_\mu \\
[f_1, P_\mu \otimes 1 +1&\otimes P_\mu] =P_\mu \otimes (a\cdot P) -a_\mu P^{\alpha}\otimes P_\alpha
\end{split}\end{equation}
 which determines the parameters $\alpha_1$ and $\alpha_2$ to be $\alpha_1=i$ and $\alpha_2=-i$, i.e.,
\begin{equation}
f_1=-i(X\cdot P)\otimes (a\cdot P) +i(a\cdot X)P_\alpha \otimes P^{\alpha}.
\end{equation}

Collecting both the first and the second order terms in the LHS of eq. \eqref{copXform}, we find
\begin{eqnarray}
\Delta X_\mu &=& X_\mu \otimes 1-X_\mu \otimes (a\cdot P)+(a\cdot X)\otimes P_\mu +X_\mu\otimes (a\cdot P)^2 -a_\mu (a\cdot X)P^{\alpha}\otimes P_\alpha \nonumber \\
&& -(a\cdot X)(a\cdot P)\otimes P_\mu -(a\cdot X)\otimes (a\cdot P) P_\mu +(a\cdot X)P_\mu \otimes (a\cdot P) -\frac{a^2}{2}X_\mu \otimes P^2 +O(a_0^3). \label{copX2}
\end{eqnarray}
We now expand $\Delta P_\mu$ to the second order and write down the most general expression for $f_2$ that is linear in $X_\mu$ and quadratic in $a_\mu$
\begin{equation}
f_2=i\beta_1(X\cdot P)\otimes (a\cdot P)^2 +i\beta_2(X\cdot P)\otimes P^2 +i\beta_3 (a\cdot X) (a\cdot P)P_\alpha\otimes P^\alpha +i\beta_4 (a\cdot X)P^2\otimes (a\cdot P) +i\beta_5 (a\cdot X)P_\alpha \otimes (a\cdot P)P^\alpha
\end{equation}
where $\beta_{1}$, ..., $\beta_{5}$ are dimensionless parameters. The parameters $\beta_i$ are determined by requiring that 
\begin{equation}
[f_2, \Delta_0 X_\mu]=-1/2 [f_1, [f_1, \Delta_0X_\mu]] + \text{second order terms from eq. \eqref{copX2}},
\end{equation}
 and an analogous expression for $[f_2, \Delta_0 P_\mu]$. This leads to
\begin{eqnarray}
f_2 &=& \frac{i}{2}(X\cdot P)\otimes (a\cdot P)^2 -\frac{ia^2}{2}(X\cdot P)\otimes P^2 -i(a\cdot X)(a\cdot P)P^\alpha \otimes P_\alpha \nonumber \\
&& +\frac{i}{2} (a\cdot X) P^2\otimes (a\cdot P) -\frac{i}{2} (a\cdot X)P_\alpha\otimes (a\cdot P)P^\alpha.
\end{eqnarray}

Repeating the procedure in the third order we find
\begin{eqnarray}
\Delta X_\mu &=& X_\mu \otimes 1-X_\mu \otimes (a\cdot P)+(a\cdot X)\otimes P_\mu +X_\mu\otimes (a\cdot P)^2 -a_\mu (a\cdot X)P^{\alpha}\otimes P_\alpha -(a\cdot X)(a\cdot P)\otimes P_\mu \nonumber \\
&& -(a\cdot X)\otimes (a\cdot P)P_\mu +(a\cdot X)P_\mu \otimes (a\cdot P) -\frac{a^2}{2}X_\mu \otimes P^2 -a^2X_\mu \otimes (a\cdot P) P^2 +X_\mu\otimes (a\cdot P)^3 \nonumber \\
&& +a^2(a\cdot X)P_\alpha P_\mu \otimes P^\alpha +a^2(a\cdot X) P^\alpha \otimes P_\alpha P_\mu +a^2(a\cdot X)\otimes P^2P_\mu +\frac{1}{2}a^2(a\cdot X)P^2 \otimes P_\mu \nonumber \\
&& -2a_\mu (a\cdot X) (a\cdot P)P^\alpha \otimes P_\alpha -a_\mu (a\cdot X)P^\alpha \otimes (a\cdot P) P_\alpha +\frac{1}{2}a_\mu (a\cdot X)P^2\otimes (a\cdot P) -(a\cdot X)\otimes (a\cdot P)^2P_\mu \nonumber \\
&& +(a\cdot X)(a\cdot P)P_\mu \otimes (a\cdot P) +(a\cdot X)P_\mu \otimes (a\cdot P)^2 -(a\cdot X) (a\cdot P) \otimes (a\cdot P)P_\mu -(a\cdot X)(a\cdot P)^2 \otimes P_\mu \nonumber\\
&& +O(a_0^4). \label{copX3}
\end{eqnarray}
We expand $\Delta P_\mu$ to the third order, then write an ansatz for $f_3$ and determine the unknown coefficients by requiring that $f_3$ satisfies 
\begin{equation}\begin{split}
[f_3, \Delta_0 X_\mu]= &-1/6[f_1, [f_1, [f_1, \Delta_0X_\mu]]] -1/2 ([f_1, [f_2, \Delta_0 X_\mu]] \\
&+[f_2, [f_1, \Delta_0 X_\mu]]) + \text{terms of order $a_0^3$ from eq. \eqref{copX3}}
\end{split}\end{equation}
 and similarly for $[f_3, \Delta_0 P_\mu]$. This gives us
\begin{eqnarray}
f_3 &=& -\frac{i}{3}(X\cdot P)\otimes (a\cdot P)^3 +\frac{i}{2}a^2 (X\cdot P)\otimes (a\cdot P)P^2 +\frac{i}{3}(a\cdot X)P_\alpha\otimes (a\cdot P)^2P^\alpha -\frac{i}{2}(a\cdot X)(a\cdot P)P^2\otimes (a\cdot P) \nonumber \\
&& -\frac{i}{2}(a\cdot X)P^2\otimes (a\cdot P)^2 +i(a\cdot X)(a\cdot P)P_\alpha \otimes (a\cdot P)P^\alpha +i(a\cdot X)(a\cdot P)^2P^\alpha \otimes P_\alpha -\frac{i}{2}a^2(a\cdot X)P^2P^\alpha\otimes P_\alpha \nonumber \\
&& -\frac{i}{2}a^2(a\cdot X)P^\alpha\otimes P^2P_\alpha -\frac{i}{2}a^2(a\cdot X)P_\alpha P_\beta \otimes P^\alpha P^\beta. 
\end{eqnarray}

The twist for the natural realization should satisfy $m\left( \mathcal{F}^{-1}\triangleright X_\mu \otimes 1 \right)=\hat{x}_\mu$. One can verify that this is satisfied in our construction. Hence, the action of
\begin{equation}
\mathcal{F}^{-1}= 1\otimes 1-f_1 -f_2 +\frac{1}{2}f_1^2+\frac{1}{2}(f_1f_2+f_2f_1)-f_3 -\frac{1}{6}f_1^3+...
\end{equation}
on $X_\mu \otimes 1$ gives $X_\mu +X_\mu (a\cdot P)+ \frac{a^2}{2} X_\mu P^2 -(a\cdot X)P_\mu+O(a^4_{0})$  , which is exactly what one gets by expanding eq. \eqref{hxnr} to the third order.
\subsection{Tensor identities for the natural realization}
A crucial point in our work are the so called tensor exchange identities (or relations) \cite{rmatrix}. The relations $\mathcal{R}$ for the natural realization can be derived by starting with the coproduct for $\hat{x}_\mu$
\begin{equation}\label{cophx}
\Delta \hat{x}_\mu= \hat{x}_\mu\otimes 1= Z^{-1}\otimes \hat{x}_\mu -a_\mu p_\alpha^{L}\otimes \hat{x}^\alpha,
\end{equation}
where $p_\mu^L=P_\mu -\frac{a_\mu}{2}\square$. Eq. \eqref{cophx} leads to
\begin{eqnarray}\label{51}
X_\mu \otimes 1 &=& \Phi_{\alpha\mu}^{-1}Z^{-1} \otimes X_\beta \Phi^{\beta\alpha} -a^\alpha \Phi_{\alpha\mu}^{-1}(P_\gamma -\frac{a_\gamma}{2}\square )\otimes X_\beta \Phi^{\beta\gamma} \nonumber \\
&=& 1\otimes X_\mu +1\otimes X_\mu (a\cdot P) -1\otimes (a\cdot X)P_\mu +P_\mu \otimes (a\cdot X)-a_\mu P^{\alpha} \otimes X_\alpha +O(a_0^2). \label{relX}
\end{eqnarray}
From eq.\eqref{relX} we can get the relations $\mathcal{R}$ expanded in the first order of the deformation parameter $a_0$. These relations can also be derived using the twist. Namely, for the symmetric algebra in $\{X_\mu\}$, denoted by $\mathcal{A}$, the coalgebra, isomorphic to $\mathcal{A}$, is given by $\Delta_0 \mathcal{A}=[(\mathcal{A}\otimes \mathcal{A})+\mathcal{I}_{0}] /\mathcal{I}_0$, where $\mathcal{I}_{0}=\mathcal{U}_{+}(\mathcal{R}_{0})(\mathcal{A}\otimes 1)\Delta_{0}\mathcal{T}$ and $\mathcal{R}_0\equiv X_\mu\otimes 1-1\otimes X_\mu\ $ are the undeformed relations that generate the equivalence class $[X_\mu\otimes 1]=[ 1\otimes X_\mu]$. If one deforms the coalgebra structure by twist, the relations also transform, into $\mathcal{R}\equiv \mathcal{F}\mathcal{R}_0\mathcal{F}^{-1}$. Now the relations $\mathcal{R}$ also induce a partition of the algebra $\mathcal{A}\otimes \mathcal{A}$ into equivalence classes. The algebra $\Delta \mathcal{A} =[(\mathcal{A}\otimes \mathcal{A})_{\mathcal{F}}+\mathcal{I}]/ \mathcal{I}$, where $\mathcal{I}=\mathcal{F}\mathcal{I}_{0}\mathcal{F}^{-1}$, together with $\Delta \mathcal{T}$ (the algebra obtained by twist deforming the coalgebra $\Delta_0 \mathcal{T}$, that is isomorphic to the symmetric algebra $\{P_\mu\}$), forms the algebra $\Delta \mathcal{H}=\Delta\mathcal{A} \Delta\mathcal{T}$, which contains all the deformed coproducts $\Delta h, \forall h\in \mathcal{H}$. It can be checked that applying our twist to the relations $\mathcal{R}_0$ we can rederive \eqref{relX}, which also shows the consistency of our construction.

Remark: $\Delta X$ and $\mathcal{F}$ are not uniquely determined. For both of them, there exists an infinite class of expressions, but all these expressions can be shown to be equivalent using the tensor identities.
\subsection{Cocycle condition}
What remains to be checked is that the twist we have constructed obeys the cocyle condition
\begin{equation}
(\mathcal{F}\otimes 1)(\Delta_0\otimes 1)\mathcal{F}= (1\otimes \mathcal{F}) (1\otimes \Delta_0)\mathcal{F}.
\end{equation}
This is trivially satisfied in the zeroth order and we have explicitly checked it for the first and the second one. 
If we write an expansion of our twist in the form
\begin{equation}
\mathcal{F}=1\otimes 1+F_1+F_2 +..., \quad F_1\equiv f_1, \quad F_2\equiv f_2+\frac{1}{2}f_1^2,
\end{equation}
then the first and the second order of the cocycle condition read
\begin{eqnarray}
F_1\otimes 1+(\Delta_0 \otimes 1)F_1 &=& 1\otimes F_1+(1\otimes \Delta_0) F_1 \label{cocycle1} \\
F_2\otimes 1+(F_1\otimes 1) (\Delta_0\otimes 1)F_1 +(\Delta_0\otimes 1)F_2 &=& 1\otimes F_2 +(1\otimes F_1) (1\otimes \Delta_0) F_1 +(1\otimes \Delta_0) F_2. \label{cocycle2}
\end{eqnarray}
The left and right hand sides of eq. \eqref{cocycle1} are
\begin{eqnarray}
LHS \eqref{cocycle1} &=& -i(X\cdot P)\otimes (a\cdot P)\otimes 1 +i(a\cdot X)P_\alpha\otimes P^\alpha\otimes 1-i (X\cdot P)\otimes 1\otimes (a\cdot P) +i(a\cdot X)P^\alpha \otimes 1\otimes P_\alpha \nonumber \\
&& -iX^\alpha\otimes P_\alpha\otimes (a\cdot P) +i(a\cdot X)\otimes P_\alpha\otimes P^\alpha \\
RHS\eqref{cocycle1} &=& -i(X\cdot P)\otimes (a\cdot P)\otimes 1 +i(a\cdot X)P_\alpha\otimes P^\alpha\otimes 1-i (X\cdot P)\otimes 1\otimes (a\cdot P) +i(a\cdot X)P^\alpha \otimes 1\otimes P_\alpha \nonumber \\
&& -i1\otimes (X\cdot P)\otimes (a\cdot P) +i1\otimes (a\cdot X) P_\alpha\otimes P^\alpha.
\end{eqnarray}
In order to compare the last two terms from the two sides, one needs to use the relations \eqref{relX} in the zeroth order ($X_\mu \otimes 1=1\otimes X_\mu +O(a_0)$). It is then easily seen that the first order of the cocycle condition is satisfied.

When the relations (\ref{51}) are used in the LHS of eq. \eqref{cocycle1}, terms of order $a_0$ in the relations (\ref{cocycle1}) give terms of order $a_0^2$ and these need to be taken into account when checking the cocycle condition in the second order.  These terms need to be added to the terms obtained by calculating the LHS of eq. \eqref{cocycle2}. A check of the cocycle condition in the second order then goes as follows: We calculate the RHS of eq. \eqref{cocycle2}, this gives 43 terms. We calculate the LHS of eq. \eqref{cocycle2}, which gives 40 terms. To the terms on the left we add the 8 second order terms that come from using the relations in the first order. Now on the left side we use the relations (in the zeroth order) and from the 40+8 terms we get precisely the 43 terms that are on the right side. Thus, we have verified that the operator we have constructed indeed satisfies all the properties of a twist up to the second order. This is in accordance with the statement that associative star product leads to cocycle condition for the twist.

\subsection{$\kappa$-Poincar\'{e} algebra}
In the natural realization, the generators of the Lorentz algebra are given by $M_{\mu\nu}= X_\mu P_\nu -X_\nu P_\mu$. We can calculate the coproduct for $M_{\mu\nu}$ using twist i.e. $\Delta M_{\mu\nu} =\mathcal{F} \Delta_0 M_{\mu\nu} \mathcal{F}^{-1}$, where $\Delta_0 M_{\mu\nu} =M_{\mu\nu} \otimes 1+1 \otimes M_{\mu\nu}$ is the primitive coproduct. Hence,
\begin{eqnarray}
\Delta M_{\mu\nu} &=& \Delta_0 M_{\mu\nu} + [f, \Delta_0 M_{\mu\nu}] +\frac{1}{2} [f,[f, \Delta_0 M_{\mu\nu}]]+... \nonumber \\
&=& \Delta_0 M_{\mu\nu} +[f_1, \Delta_0 M_{\mu\nu}] +O(a_0^2) \nonumber \\
&=& \Delta_0 M_{\mu\nu} -(X\cdot P) a_\mu P_\nu -a_\nu X_\mu P^\alpha \otimes P_\alpha +(X\cdot P)\otimes a_\nu P_\mu +a_\mu X_\nu P^\alpha \otimes P_\alpha +O(a_0^2) \nonumber \\
&=& \Delta_0 M_{\mu\nu} -a_\mu P^\alpha \otimes M_{\alpha \nu} +a_\nu P^\alpha \otimes M_{\alpha \mu} +O(a_0^2), \label{deltaM}
\end{eqnarray}
which is what one gets when expanding the known expression for $\Delta M_{\mu\nu}$ in natural realization \cite{Meljanac-3, ms06, ms11}. It is easily checked that our twist also gives the correct expressions in higher orders. However, we emphasise that the tensor identities play a crucial role in perturbative calculation order by order. In going from the third to the fourth line in eq. \eqref{deltaM}, we have used the identities in the zeroth order, which generate terms of order $a_0^2$, which have to be taken into account when calculating $\Delta M$ in the second order (and analogously for higher orders). We point out that twist $\mathcal{F}$ for natural realization can not be written in terms of Poincar\'{e} generators, however coalgebra of Poincar\'{e} algebra is expressed in terms of Poincar\'{e} generators.

\subsection{$R$-matrix}
The $R$-matrix\footnote{Quantum universal $R$-matrix is defined as a solution of the Yang-Baxter equation \cite{kassel}.}, $R:\Delta \mathcal{H}\mapsto\tilde{\Delta}\mathcal{H}$, satisfies
\begin{equation}
R\Delta h R^{-1}=\tilde{\Delta}h, \ \ \forall h\in\mathcal{H},
\end{equation}
where $\tilde{\Delta}h=\tau_{0}\Delta h\tau_{0}$ is the opposite coproduct and  $\tau_0$ is the flip operator, $\tau_0 (h_1\otimes h_2)= h_2 \otimes h_1, \forall h_1, h_2 \in H$.
From the twist for the natural realization, we can construct the $R$-matrix for the $\kappa$-Poincar\'{e} algebra. With $\tilde{\mathcal{F}}$ defined by $\tilde{\mathcal{F}}=\tau_0\mathcal{F}\tau_0$, the $R$-matrix is given by
\begin{eqnarray}\label{rmatrix}
R &=& \tilde{\mathcal{F}} \mathcal{F}^{-1} =e^{\tilde{f}_1 +\tilde{f}_2+ ...} e^{-f_1 -f_2- ...} =e^{\tilde{f}_1 -f_1 +\tilde{f}_2 -f_2 -\frac{1}{2} [\tilde{f}_1, f_1] +...} \nonumber \\
&=& \exp \Biggl( i(X\cdot P)\otimes (a\cdot P) -i(a\cdot  P)\otimes (X\cdot P) -i(a\cdot X) P^\alpha \otimes P_\alpha +i P^\alpha \otimes (a\cdot X)P_\alpha \nonumber \\
&& +\frac{i}{2} (X\cdot P)\otimes (a\cdot P)^2 -\frac{i}{2} (a\cdot P)^2 \otimes (X\cdot P) -\frac{i}{2} a^2(X\cdot P)\otimes P^2 +\frac{i}{2} a^2 P^2\otimes (X\cdot P) \nonumber \\
&& -i(a\cdot X)(a\cdot P) P^\alpha \otimes P_\alpha +iP^\alpha \otimes (a\cdot X)(a\cdot P)P_\alpha +\frac{i}{2}(a\cdot X)P^2\otimes (a\cdot P) -\frac{i}{2} (a\cdot P)\otimes (a\cdot X)P^2 \nonumber \\ 
&& -\frac{i}{2} (a\cdot X)P_\alpha \otimes (a\cdot P)P^\alpha +\frac{i}{2} (a\cdot P)P^\alpha \otimes (a\cdot X)P_\alpha +\frac{i}{2} (X\cdot P)(a\cdot P)\otimes (a\cdot P) -\frac{i}{2} (a\cdot P)\otimes (X\cdot P)( a\cdot P) \nonumber \\
&& -\frac{i}{2} a^2(X\cdot P)P_\alpha \otimes P^\alpha +\frac{i}{2} a^2P^\alpha \otimes (X\cdot P)P_\alpha +O(a_0^3) \Biggr).
\end{eqnarray}
In ref. \cite{Govindarajan-2}, the authors have given the universal $R$-matrix for the $\kappa$-Poincar\'{e} algebra in terms of a family of realizations given by $\hat{x}_i =x_iZ^{-\lambda}, \hat{x}_0 =x_0 -a_0(1-\lambda) x_kp_k$, where $\lambda$ is a real parameter. This family of realizations is related to the natural realization by similarity transformations. Expanding $R$-matrix (\ref{rmatrix}) and using $X_\mu=x_\mu +O(a_0)$ and  $P_\mu=p_\mu +O(a_0)$ we find that
\begin{eqnarray}
R &=& 1\otimes 1+ iX\cdot P\otimes (a\cdot P)-i(a\cdot P)\otimes (X\cdot P) +O(a_0^2) = 1\otimes 1 -iX_kP_k \otimes a_0P_0 +ia_0P_0 \otimes X_kP_k +O(a_0^2) \nonumber \\
&=& 1\otimes 1 -ix_kp_k \otimes a_0p_0 +ia_0p_0 \otimes x_kp_k +O(a_0^2),
\end{eqnarray}
which agrees with the result from ref. \cite{rmatrix}. It can be easily checked that the results also agree in higher orders. Note that, although the twist $\mathcal{F}$ can not be expressed in terms of $\kappa$-Poincar\'{e} generators, the $R$-matrix \cite{rmatrix} is expressed in terms of $\kappa$-Poincar\'{e} generators up to the fifth order \cite{young}.

\section{Different realizations and similarity transformations}
\subsection{Quantum phase space}
Different realizations correspond to different bases of undeformed Heisenberg algebras. Undeformed Heisenberg algebras are mathematical framework for quantum phase space. Different bases of Heisenberg algebras are connected via quantum canonical transformations. We are looking for a special set of these transformations, i.e. similarity transformations, which connect two different bases of Heisenberg algebras, algebra $\mathcal{H}^{(1)}$ generated by $\left\{ x^{(1)}, p^{(1)} \right\}$ and  algebra $\mathcal{H}^{(2)}$ generated by $\left\{x^{(2)},p^{(2)}\right\}$:
\begin{equation}\begin{split}\label{similarity}
&x^{(2)}_{\mu}={\cal S}^{(1,2)} x^{(1)}_{\mu} \left[{\cal S}^{(1,2)}\right]^{-1}\equiv x^{(1)}_{\alpha}\left[\psi^{(1,2)}\right]^{\alpha}_{\ \mu} \\
&p^{(2)}_{\mu}={\cal S}^{(1,2)} p^{(1)}_{\mu} \left[{\cal S}^{(1,2)}\right]^{-1}\equiv \Lambda^{(1,2)}_{\mu} \\
\end{split}\end{equation}
and inversely
\begin{equation}\begin{split}\label{similarity1}
&x^{(1)}_{\mu}={\cal S}^{(2,1)} x^{(2)}_{\mu} \left[{\cal S}^{(2,1)}\right]^{-1}\equiv x^{(2)}_{\alpha}\left[\psi^{(2,1)}\right]^{\alpha}_{\ \mu} \\
&p^{(1)}_{\mu}={\cal S}^{(2,1)} p^{(2)}_{\mu} \left[{\cal S}^{(2,1)}\right]^{-1}\equiv \Lambda^{(2,1)}_{\mu} \\
\end{split}\end{equation}
where $\psi^{(1,2)}$ and $\Lambda^{(1,2)}_{\mu}$ are  functions of $p^{(1)}$ only, $\psi^{(2,1)}$ and $\Lambda^{(2,1)}_{\mu}$ are  functions of $p^{(2)}$ only, ${\cal S}^{(1,2)}$ is a function of $x^{(1)}_{\mu}$ and $p^{(1)}_{\mu}$, and ${\cal S}^{(2,1)}$ is a function of $x^{(2)}_{\mu}$ and $p^{(2)}_{\mu}$. Equations (\ref{similarity}) and (\ref{similarity1}) imply
\begin{equation}\begin{split}\label{veze}
\mathcal{S}^{(2,1)}(p^{(2)})&=\left[\mathcal{S}^{(1,2)}(p^{(1)})\right]^{-1}\\
\psi^{(2,1)}(p^{(2)})&=\left[\psi^{(1,2)}(p^{(1)})\right]^{-1}\\
\Lambda^{(2,1)}(p^{(2)})&=\Lambda^{(2,1)}\left(\Lambda^{(1,2)}(p^{(1)})\right)
\end{split}\end{equation} 
Since both $\left\{x^{(1)},p^{(1)}\right\}$ and $\left\{x^{(2)},p^{(2)}\right\}$  generate the undeformed Heisenberg algebras for each of them, eq. (\ref{H}) holds and together with (\ref{similarity}, \ref{similarity1}) we have:
\begin{equation}\label{uvet}
\left[\psi^{(1,2)}\right]^{-1}_{\mu\nu}=\frac{\partial \Lambda^{(1,2)}_{\mu}}{\partial (p^{(1)})^{\nu}} \ , \quad \left[\psi^{(2,1)}\right]^{-1}_{\mu\nu}=\frac{\partial \Lambda^{(2,1)}_{\mu}}{\partial (p^{(2)})^{\nu}}
\end{equation}
Equations (\ref{similarity}, \ref{similarity1}, \ref{veze}, \ref{uvet} ) can be written in a unified way:
\begin{subequations}\label{veze2}
\begin{equation}
x^{(j)}_{\mu}={\cal S}^{(i,j)} x^{(i)}_{\mu} \left[{\cal S}^{(i,j)}\right]^{-1}\equiv x^{(i)}_{\alpha}\left[\psi^{(i,j)}\right]^{\alpha}_{\ \mu}
\end{equation}
\begin{equation}
p^{(j)}_{\mu}={\cal S}^{(i,j)} p^{(i)}_{\mu} \left[{\cal S}^{(i,j)}\right]^{-1}\equiv \Lambda^{(i,j)}_{\mu}
\end{equation}
\begin{equation}
\left[\psi^{(i,j)}\right]^{\alpha}_{\ \beta}\equiv\left[\psi^{(i,j)}\right]^{\alpha}_{\ \beta}(p^{i}),\ \quad \Lambda^{(i,j)}_{\mu}\equiv\Lambda^{(i,j)}_{\mu}(p^{(i)}),\ \quad \ \mathcal{S}^{(i,j)}\equiv \mathcal{S}^{(i,j)}(x^{(i)},p^{(i)})=\left[\mathcal{S}^{(j,i)}\right]^{-1}(x^{(j)},p^{(j)})
\end{equation}
\begin{equation}
\left[\psi^{(i,j)}\right]^{-1}_{\mu\nu}=\frac{\partial \Lambda^{(i,j)}_{\mu}}{\partial (p^{(i)})^{\nu}}
\end{equation}
\end{subequations}
where $i, j={1,2}$ and  $i\neq j$.
For these special transformations, $\mathcal{S}^{(i,j)}$ can be written as $\mathcal{S}^{(i,j)}=\text{exp}(x^{(i)}_{\alpha}\Sigma^{\alpha})$, where $\Sigma^{\alpha}$ is  a function of $p^{(i)}$ . Then we have 
\begin{equation}
p^{(j)}_{\mu}=p^{(i)}_{\mu}+\frac{\text{e}^{O}-1}{O}\Sigma_{\mu} , \quad i,j=1,2\ \text{and} \ i\neq j
\end{equation}
where $O=\Sigma^{\alpha}\frac{\partial}{\partial p^{(i)}_{\alpha}}$.

 Undeformed Heisenberg algebras $\mathcal{H}^{(1)}$ and $\mathcal{H}^{(2)}$ are isomorphic $\mathcal{H}^{(1)}\cong \mathcal{H}^{(2)}$. We can define actions\footnote{algebra $\mathcal{A}^{(1)}$ is generated by $x^{(1)}_{\mu}$ and algebra $\mathcal{A}^{(2)}$ by $x^{(2)}_{\mu}$.} $\triangleright_{(1)}$ and $\triangleright_{(2)}$  as mapings $\ \triangleright_{(1)}: \mathcal{H}^{(1)}\mapsto \mathcal{A}^{(1)}$ and $\ \triangleright_{(2)}: \mathcal{H}^{(2)}\mapsto \mathcal{A}^{(2)}$  by
\begin{equation}\begin{split}
&x^{(1)}_{\mu}\triangleright_{(1)} 1=x^{(1)}_{\mu}, \quad x^{(2)}_{\mu}\triangleright_{(2)}  1=x^{(2)}_{\mu},\\
&\ \ p^{(1)}_{\mu}\triangleright_{(1)} 1=0, \quad p^{(2)}_{\mu} \triangleright_{(2)} 1=0,\\
&f^{(2)}\triangleright_{(1)} 1= f^{(1)}, \quad f^{(1)}\triangleright_{(2)} 1=f^{(2)},
\end{split}\end{equation}
where $f^{(1)}\equiv f^{(1)}(x^{(1)})$ and $f^{(2)}\equiv f^{(2)}(x^{(2)})$ are elements of $\mathcal{A}^{(1)}$ and $\mathcal{A}^{(2)}$ respectively. The star products $*_{(1)}$ and $*_{(2)}$ are defined by
\begin{subequations}\label{startildestar}
\begin{equation}
f^{(2)} g^{(2)}\triangleright_{(1)} 1= f^{(1)} *_{(1)} g^{(1)}
\end{equation}
\begin{equation}
f^{(1)} g^{(1)} \triangleright_{(2)}  1=f^{(2)} *_{(2)} g^{(2)}
\end{equation}
\end{subequations}
and they are both commutative and associative. If $\Lambda^{(i,j)}_{\mu}(p^{(i)})\neq p^{(i)}_{\mu}$ (that is $\Lambda^{(i,j)}$ has higher powers in $p^{(i)}$) then the star products $*_{(1)}$ and $*_{(2)}$ are nonlocal.

For algebra $\mathcal{H}^{(1)}$ we can construct the coproduct $\Delta^{(1)}_{0}$, and for algebra $\mathcal{H}^{(2)}$ the coproduct $\Delta^{(2)}_{0}$ that is
\begin{equation}\label{coproducts}
\Delta^{(i)}_{0}x^{(i)}_{\mu}=x^{(i)}_{\mu}\otimes 1, \quad \Delta^{(i)}_{0}p^{(i)}_{\mu}=p^{(i)}_{\mu}\otimes 1+1\otimes p^{(i)}_{\mu}
\end{equation}
where we generated an equivalence class in $\mathcal{H}^{(i)}\otimes \mathcal{H}^{(i)}$ by the ideal $\mathcal{I}^{(i)}_{0}=\mathcal{U}_{+}(\mathcal{R}^{(i)}_{0})(\mathcal{A}^{(i)}\otimes 1)\Delta^{(i)}_{0}\mathcal{T}^{(i)}$, where $\mathcal{R}^{(i)}_{0}\equiv x^{(i)}_{\mu}\otimes 1 - 1\otimes x^{(i)}_{\mu}$. It can be shown that $\Delta^{(i)}_{0}\left(\mathcal{A}^{(i)}\right)=\left(\mathcal{A}^{(i)}\otimes 1 + \mathcal{I}^{(i)}_{0}\right) / \mathcal{I}^{(i)}_{0}=\left(\mathcal{A}^{(i)}\otimes \mathcal{A}^{(i)}+\mathcal{I}^{(i)}_{0}\right) / \mathcal{I}^{(i)}_{0}$ is an algebra isomorphic to $\mathcal{A}^{(i)}$.
Using (\ref{veze2}) and  homomorphism of coproduct $\Delta^{(i)}_{0}(h_{1}h_{2})=\Delta^{(i)}_{0}(h_{1})\Delta^{(i)}_{0}(h_{2})$ for every $h_{1},h_{2}\in \mathcal{H}^{(i)}$ we have
\begin{equation}\begin{split}\label{veze3}
\Delta^{(i)}_{0}x^{(i)}_{\mu}&=\left(\mathcal{S}^{(j,i)}\otimes\mathcal{S}^{(j,i)}\right)\Delta^{(j)}_{0}x^{(j)}_{\mu}\left(\mathcal{S}^{(i,j)}\otimes\mathcal{S}^{(i,j)}\right)\\
\Delta^{(i)}_{0}p^{(i)}_{\mu}&=\left(\mathcal{S}^{(j,i)}\otimes\mathcal{S}^{(j,i)}\right)\Delta^{(j)}_{0}p^{(j)}_{\mu}\left(\mathcal{S}^{(i,j)}\otimes\mathcal{S}^{(i,j)}\right)\\
\end{split}\end{equation}
and also
\begin{equation}\begin{split}
\Delta^{(i)}_{0}x^{(j)}_{\mu}&=\Delta^{(i)}_{0}\left(x^{(j)}_{\alpha}\left[\psi^{(i,j)}\right]^{\alpha}_{\ \mu}\right)=\Delta^{(i)}_{0}\left(x^{(i)}_{\alpha}\right)\Delta^{(i)}_{0}\left(\left[\psi^{(i,j)}\right]^{\alpha}_{\ \mu}\right)\neq x^{(j)}_{\mu}\otimes 1\\
\Delta^{(i)}_{0}p^{(j)}_{\mu}&=\Delta^{(i)}_{0}\left(\mathcal{S}^{(i,j)}\right)\Delta^{(i)}_{0}\left(p^{(i)}_{\mu}\right)\Delta^{(i)}_{0}\left(\left[\mathcal{S}^{(i,j)}\right]^{-1}\right)\\
&=\Delta^{(i)}_{0}\left(\mathcal{S}^{(i,j)}\right)p^{(i)}_{\mu}\otimes \Delta^{(i)}_{0}\left(\left[\mathcal{S}^{(i,j)}\right]^{-1}\right)+\Delta^{(i)}_{0}\left(\mathcal{S}^{(i,j)}\right)\otimes p^{(i)}_{\mu}\Delta^{(i)}_{0}\left(\left[\mathcal{S}^{(i,j)}\right]^{-1}\right)\\
&\neq p^{(j)}_{\mu}\otimes 1+ 1\otimes p^{(j)}_{\mu}
\end{split}\end{equation}
where $i,j=1,2$ and $i\neq j$.
Note that the coproduct $\Delta_{0}$ is a unique mathematical object and $\Delta^{(i)}_{0}$ is its realization in algebra $\mathcal{H}^{(i)}$.
Since the star products in (\ref{startildestar}) are commutative and associative, both $\Delta^{(1)}_{0}$ and $\Delta^{(2)}_{0}$ are cocommutative (\ref{coco}) and coassociative (\ref{coass}) 
\begin{subequations}
\begin{equation}\label{coco}
\tilde{\Delta}^{(i)}_{0}h=\tau_{0}\ \Delta^{(i)}_{0}h\ \tau_{0}=\Delta^{(i)}_{0} h,
\end{equation}
\begin{equation}\label{coass}
\left(\Delta^{(i)}_{0}\otimes 1\right)\Delta^{(i)}_{0}=\left(1\otimes\Delta^{(i)}_{0}\right)\Delta^{(i)}_{0},
\end{equation}
\end{subequations}
for any $h\in \mathcal{H}^{(i)}$, where $\tau_{0}$ is the flip operator defined by $\tau_{0}:\ h_1 \otimes h_2 \mapsto h_2 \otimes h_1, \forall h_{1},h_{2}\in \mathcal{H}^{(i)}$. Cocomutativity of the coproduct leads to trivial $R$-matrix, i.e. $R=1\otimes 1$ (up to the right ideal $\mathcal{J}_{0}$).

The relation between coproducts $\Delta^{(1)}_{0}$ and $\Delta^{(2)}_{0}$ defines the bidifferential operators $\mathcal{F}^{(1,2)}$ and $\mathcal{F}^{(2,1)}$ via 
 \begin{subequations}\label{ffx}
 \begin{equation}
 \Delta^{(2)}_{0}h=\mathcal{F}^{(1,2)}\left(\Delta^{(1)}_{0}h\right)\left[\mathcal{F}^{(1,2)}\right]^{-1},
 \end{equation}
 \begin{equation}
   \Delta^{(1)}_{0}h=\mathcal{F}^{(2,1)}\left(\Delta^{(2)}_{0}h\right)\left[\mathcal{F}^{(2,1)}\right]^{-1},
 \end{equation}
 \end{subequations}
where $h\in \mathcal{H}^{(i)}$ and $\mathcal{F}^{(i,j)}\equiv \mathcal{F}^{(i,j)}(x^{(i)},p^{(i)})$.  From (\ref{ffx}) we get
\begin{equation}
\mathcal{F}^{(1,2)}(x^{(1)},p^{(1)})=\left[\mathcal{F}^{(2,1)}\right]^{-1}(x^{(2)},p^{(2)})
\end{equation}
or in a unified way
\begin{equation}\begin{split}
&\Delta^{(j)}_{0}h=\mathcal{F}^{(i,j)}\left(\Delta^{(i)}_{0}h\right)\left[\mathcal{F}^{(i,j)}\right]^{-1}\\
\mathcal{F}^{(i,j)}&(x^{(i)},p^{(i)})=\left[\mathcal{F}^{(j,i)}\right]^{-1}(x^{(j)},p^{(j)})
\end{split}\end{equation}
where $i, j={1,2}$ and  $i\neq j$.
Operators $\mathcal{F}^{(1,2)}$ and $\mathcal{F}^{(2,1)}$ satisfy all the properties of a twist, that is the cocycle (\ref{CO}) and the normalization (\ref{NO}) condition. Star products in (\ref{startildestar}) can also be defined using the twist
\begin{equation}
f^{(i)} *_{(i)} g^{(i)}=m\left(\mathcal{F}^{(j,i)}\triangleright_{(i)} f^{(i)}\otimes g^{(i)}\right)
\end{equation}
Note that
\begin{equation}
x^{(j)}_{\mu}=m\left(\mathcal{F}^{(j,i)}\triangleright_{(i)}\left(x^{(i)}_{\mu}\otimes 1\right)\right)=x^{(i)}_{\alpha}\left[\psi^{(i,j)}\right]^{\alpha}_{\ \mu}
\end{equation}
For the relations $\mathcal{R}^{(1)}_{0}$ and $\mathcal{R}^{(2)}_{0}$ we have
\begin{equation}
\mathcal{R}^{(j)}_{0}=\mathcal{F}^{(i,j)}\mathcal{R}^{(i)}_{0}\mathcal{F}^{(j,i)}
\end{equation}
The coproduct and twist can also be considered as mappings 
\begin{equation}\begin{split}
\Delta^{(i)}_{0}h&:\Delta^{(i)}_{0}\mathcal{H}^{(i)}\mapsto\Delta^{(i)}_{0}\mathcal{H}^{(i)}\\
\mathcal{F}^{(i,j)}&: \Delta^{(i)}_{0}\mathcal{H}^{(i)}\mapsto\Delta^{(j)}_{0}\mathcal{H}^{(j)}
\end{split}\end{equation}
where $\Delta^{(i)}_{0}\mathcal{H}^{(i)}=\left((\mathcal{A}^{(i)}\otimes1)\Delta^{(i)}_{0}\mathcal{T}^{(i)}+\mathcal{I}^{(i)}_{0}\right)/\mathcal{I}^{(i)}_{0}$.
Using (\ref{veze2}, \ref{coproducts}, \ref{veze3}, \ref{ffx}) we can relate the twist $\mathcal{F}^{(i,j)}$ with the similarity transformation $\mathcal{S}^{(i,j)}$
\begin{equation}\begin{split}\label{prekos}
\Delta^{(j)}_{0}p^{(i)}_{\mu}&=\mathcal{F}^{(i,j)}\Delta^{(i)}_{0}p^{(i)}_{\mu}\mathcal{F}^{(j,i)}\\
&=\mathcal{F}^{(i,j)}\left(\mathcal{S}^{(j,i)}\otimes\mathcal{S}^{(j,i)}\right)\Delta^{(j)}_{0}p^{(j)}_{\mu}\left(\mathcal{S}^{(i,j)}\otimes\mathcal{S}^{(i,j)}\right)\mathcal{F}^{(j,i)}\\
\text{(l.h.s.)}&=\left(\Delta^{(j)}_{0}\mathcal{S}^{(j,i)}\right)\Delta^{(j)}_{0}p^{(j)}_{\mu}\left(\Delta^{(j)}_{0}\mathcal{S}^{(i,j)}\right)
\end{split}\end{equation}
and from last two lines it follows
\begin{equation}
\mathcal{F}^{(i,j)}=\Delta^{(j)}_{0}\mathcal{S}^{(j,i)}\left(\mathcal{S}^{(i,j)}\otimes\mathcal{S}^{(i,j)}\right)
\end{equation}
So far  we have completely defined connection between two quantum phase spaces, that is undeformed Heisenberg algebras $\mathcal{H}^{(1)}$ and $\mathcal{H}^{(2)}$ and illustrated all the mathematical tools (coproduct, twist, star product) which are usually used in the noncommutative setting.

\subsection{$\kappa$-deformed phase space in  two different  realizations}
Now we want to analyze two different realizations of NC space (\ref{kappa}), and illustrate the procedure of connecting the twist operator (\ref{cotw}) in one realization with the twist operator in another realization.
The NC coordinates $\hat{x}$ can be realized in terms of the algebra $\mathcal{H}^{(1)}$ and also in terms of algebra $\mathcal{H}^{(2)}$. 
For two different realizations of NC coordinates we can write
\begin{equation}
\hat{x}_{\mu}=x^{(1)}_{\alpha}\left[\varphi^{(1)}\right]^{\alpha}_{\ \mu}=x^{(2)}_{\alpha}\left[\varphi^{(2)}\right]^{\alpha}_{\ \mu},
\end{equation}
where $\left[\varphi^{(i)}\right]^{\alpha}_{\ \mu}\equiv\left[\varphi^{(i)}\right]^{\alpha}_{\ \mu}(p^{(i)})$ is a function of $p^{(i)}$ only.  Since both $\left\{x^{(1)},p^{(1)}\right\}$ and $\left\{x^{(2)},p^{(2)}\right\}$  generate the undeformed Heisenberg algebra, they can be connected via similarity transformations of the form (\ref{similarity}) and we get
\begin{equation}
\left[\varphi^{(i)}\right]^{\alpha}_{\ \beta}=\left[\psi^{(i,j)}\right]^{\alpha}_{\ \sigma}\left[\varphi^{(j)}\right]^{\sigma}_{\ \beta}
\end{equation}
where $i, j={1,2}$ and  $i\neq j$. We can introduce the deformed twist $\mathcal{F}^{(i,\varphi^{(i)})}$ like in (\ref{cotw}) which will depend on the realization $\varphi^{(i)}(p^{(i)})$. Twist $\mathcal{F}^{(i,\varphi^{(i)})}$ can be calculated for a given realization $\varphi^{(i)}(p^{(i)})$ as described in section II(see (\ref{D}-\ref{formule})). Hence, for every $h\in \mathcal{H}^{(i)}$  we have
\begin{equation}\begin{split}\label{cotwbar}
&\Delta h=\mathcal{F}^{(i,\varphi^{(i)})} \Delta^{(i)}_{0} h\left[\mathcal{F}^{(i,\varphi^{(i)})}\right]^{-1} \\
\hat{f}(\hat{x})\hat{g}(\hat{x})\triangleright_{(i)} 1&=f^{(i)}\star_{\varphi^{(i)}} g^{(i)}=m\left(\left[\mathcal{F}^{(i,\varphi^{(i)})}\right]^{-1}\triangleright_{(i)}(f^{(i)}\otimes g^{(i)})\right)
\end{split}\end{equation}
where it must be pointed out that the deformed coproduct $\Delta$ and star product $\star_{\varphi^{(i)}}$ depend on the choice of realization $\varphi^{(i)}$.
Now we can relate twists in different realizations $\mathcal{F}^{(1,\varphi^{(1)})}$ and $\mathcal{F}^{(2,\varphi^{(2)})}$. Using (\ref{similarity}, \ref{ffx}, \ref{cotwbar}) we have 
\begin{equation}\begin{split}
\Delta h&=\mathcal{F}^{(i,\varphi^{(i)})}\left(\Delta^{(i)}_{0}h\right) \left[\mathcal{F}^{(i,\varphi^{(i)})}\right]^{-1}\\
&=\mathcal{F}^{(i,\varphi^{(i)})}\left(\mathcal{F}^{(j,i)}\left(\Delta^{(j)}_{0}h\right)\mathcal{F}^{(i,j)}\right) \left[\mathcal{F}^{(i,\varphi^{(i)})}\right]^{-1}\\
\text{(l.h.s.)}&\equiv\mathcal{F}^{(j,\varphi^{(j)})}\left(\Delta^{(j)}_{0}h\right) \left[\mathcal{F}^{(j,\varphi^{(j)})}\right]^{-1}
\end{split}\end{equation}
which leads to 
\begin{equation}\label{vezatwistova}
\mathcal{F}^{(j,\varphi^{(j)})}=\mathcal{F}^{(i,\varphi^{(i)})}\mathcal{F}^{(j,i)}
\end{equation}
Note that the composition of the twists in (\ref{vezatwistova}) is  also a twist. Since, $\mathcal{F}^{(i,j)}$ could be expressed via similarity transformations (\ref{prekos}), eq.(\ref{vezatwistova}) states that if we have a twist in one realization and we know the connection between this realization with  another one, then we can easily get twist in  this other realization. The procedure presented in this section implies that if there exists a twist in one realization , then it exists in any other realization which can be related with the similarity transformations. The results of this section will be demonstrated by two explicit examples, where in the first one we look at relation between natural and left covariant realization \cite{Meljanac-3} and in the second one the relation between left covariant and left noncovariant realization (corresponds to the left ordering \cite{Meljanac-4, ms06}).

\section{Examples}
\subsection{Natural and left covariant}
Here we demonstrate how to get the twist for the natural realization from the known twist for the so called left covariant realization and the relative twist calculated using the similarity transformations between the two undeformed Heisenberg algebras. We will denote $\left\{x^{(1)},p^{(1)}\right\}\equiv \left\{X,P\right\}$ as generators in natural realization and $\left\{x^{(2)},p^{(2)}\right\}\equiv \left\{x^L,p^L\right\}$ as generators in left covariant realization. The left covariant realization of the $\kappa$-Minkowski spacetime is given by $\hat{x}_\mu=x_\mu^L (1+a\cdot p^L)$, where $\{x_\mu^L\}$ are commutative spacetime coordinates that, together with the associated momenta $\{p_\mu^L\}$ ($[p_\mu^L, x_\nu^L]=-i\eta_{\mu\nu}$), generate an undeformed Heisenberg algebra. The relations between this Heisenberg algebra and the one generated by $\{X_\mu,P_\mu\}$ are given by
\begin{eqnarray}
P_\mu &=& p_\mu^L -\frac{a_\mu}{2}(p^L)^2 \frac{1}{1+(a\cdot p^L)} \label{PpL} \\
X_\mu &=& x_\mu^L +(a\cdot x^L)\frac{p_\mu^L+\frac{a_\mu}{2}\square}{1-\frac{a^2}{2}\square} =x^{L \alpha} \chi_{\alpha\mu}^{-1}(p^L) \label{XxL} \\
p_\mu^L &=& P_\mu -\frac{a_\mu}{2}\frac{1-\sqrt{1+a^2P^2}}{a^2/2} \label{pLP} \\
x_\mu^L &=& X_\mu -\frac{(a\cdot X)P_\mu}{\sqrt{1+a^2P^2}+(a\cdot P)} =X^\alpha \chi_{\alpha\mu}(P). \label{xLX}
\end{eqnarray}

We now define the coproduct $\Delta_0'$ by
\begin{equation}
\Delta_0' p_\mu^L \equiv p_\mu^L\otimes 1 +1\otimes p_\mu^L, \quad \Delta_0' x_\mu^L \equiv x_\mu^L \otimes 1,
\end{equation}
and we calculate $\Delta_0'$ for $X_\mu$ and $P_\mu$ using eqs. \eqref{PpL}-\eqref{XxL} and the homomorphism property of the coproduct. Then we express the result  in terms of $X_\mu$ and $P_\mu$, using eqs. \eqref{pLP} and \eqref{xLX}
\begin{eqnarray}
\Delta_0' P_\mu &=& \Delta_0' \left(p_\mu^L -\frac{a_\mu}{2}(p^L)^2 \frac{1}{1+(a\cdot p^L)} \right) \nonumber \\
&=& P_\mu\otimes 1 +1\otimes P_\mu -a_\mu P_\alpha\otimes P^\alpha +a_\mu ((a\cdot P)P_\alpha \otimes P^\alpha +P_\alpha \otimes P^\alpha (a\cdot P)) -\frac{a_\mu}{2}a^2 (P^2P_\alpha \otimes P^\alpha +P_\alpha \otimes P^\alpha P^2) \nonumber \\
&& -\frac{a_\mu}{4}a^2P^2\otimes P^2 +a_\mu \bigl((a\cdot P)^2P_\alpha \otimes P^\alpha +P_\alpha \otimes P^\alpha (a\cdot P)^2 +2(a\cdot P) P_\alpha\otimes P^\alpha (a\cdot P) \bigr)+ O(a_0^4) \\
\Delta_0' X_\mu &=& (X_\alpha \chi^{\alpha\beta}(P)\otimes 1)(\Delta_0' \chi_{\beta\mu}^{-1}(P)) \nonumber \\
&=& X_\mu \otimes 1+(a\cdot X)\otimes P_\mu -(a\cdot X)(a\cdot P)\otimes P_\mu -a_\mu (a\cdot X)P_\alpha\otimes P^\alpha -2a_\mu (a\cdot X)(a\cdot P)P_\alpha \otimes P^\alpha \nonumber \\
&& -a_\mu (a\cdot X)P^\alpha \otimes (a\cdot P)P_\alpha -(a\cdot X) (a\cdot P)^2 \otimes P_\mu +\frac{a^2}{2}(a\cdot X)\otimes 1 \Bigl( 1\otimes P^2P_\mu +P_\mu \otimes P^2 +P^2\otimes P_\mu \nonumber \\
&& +2P^\alpha P_\mu \otimes P_\alpha +2P_\alpha \otimes P^\alpha P_\mu \Bigr) +O(a_0^4)
\end{eqnarray}
Our next step is to find the relative twist $\mathcal{F}_{P,p^L}$ such that
\begin{equation}
\Delta_0' X_\mu =\mathcal{F}_{P,p^L} \Delta_0 X_\mu \mathcal{F}_{P,p^L}^{-1}, \quad \Delta_0'P_\mu = \mathcal{F}_{P,p^L} \Delta_0 P_\mu \mathcal{F}_{P,p^L}^{-1},
\end{equation}
where $\Delta_0 X_\mu =X_\mu \otimes 1$ and $\Delta_0 P_\mu =P_\mu \otimes 1+1 \otimes P_\mu$. Using the pertubative method outlined in section \ref{2}, we find
\begin{eqnarray}
\mathcal{F}_{P,p^L} &=& \exp \Bigl( i(a\cdot X)P^\alpha \otimes P_\alpha +i(a\cdot X) (a\cdot P)P^\alpha \otimes P_\alpha +\frac{i}{2}(a\cdot X)P^\alpha (a\cdot P)P_\alpha +i(a\cdot X)(a\cdot P)^2P^\alpha \otimes P_\alpha \nonumber \\
&& +i(a\cdot X)(a\cdot P)P^\alpha \otimes (a\cdot P)P_\alpha +\frac{i}{3} (a\cdot X)P_\alpha \otimes (a\cdot P)^2P^\alpha -\frac{i}{2} a^2(a\cdot X)P^\alpha \otimes P^2P_\alpha -\frac{i}{4} a^2(a\cdot X)P^2 \otimes P^2 \nonumber \\
&& -\frac{i}{2} a^2(a\cdot X)P^2 P_\alpha \otimes P^\alpha -\frac{i}{2}a^2 (a\cdot X)P_\alpha P_\beta \otimes P^\alpha P^\beta +O(a_0^4) \Bigr).
\end{eqnarray}
The twist for the left covariant realization, which is a special example of a Jordanian twist, is known to be
\begin{equation}
\mathcal{F}_L =\exp \bigl( -i(x^L \cdot p^L) \otimes \ln (1+(a\cdot p^L)) \bigr).
\end{equation}
Using eqs. \eqref{pLP} and \eqref{xLX} this can be written in terms of $X_\mu$ and $P_\mu$
\begin{eqnarray}
\mathcal{F}_L &=& \exp \Bigl( -i(X\cdot P)\otimes (a\cdot P) +\frac{i}{2}a^2 (X\cdot P)\otimes P^2 -\frac{i}{2}(X\cdot P)\otimes (a\cdot P)^2 -\frac{i}{2} (a\cdot X)P^2 \otimes a\cdot P \nonumber \\
&& +\frac{i}{2}a^2 (X\cdot P) \otimes (a\cdot P)P^2 -\frac{i}{3}(X\cdot P)\otimes (a\cdot P)^3 +\frac{i}{4} a^2(a\cdot X)P^2\otimes P^2 \nonumber \\
&& -\frac{i}{4}(a\cdot X)P^2 \otimes (a\cdot P)^2 -\frac{i}{2}(a\cdot X)(a\cdot P)P^2 \otimes (a\cdot P )+O(a_0^4) \Bigr).
\end{eqnarray}
With $\mathcal{F}_{P,p^L} =\exp(f_{P,p^L})$ and $\mathcal{F}_L =\exp(f_L)$, we now calculate the twist $\mathcal{F}$ for the natural realization as
\begin{eqnarray}
\mathcal{F} &=& \mathcal{F}_L \mathcal{F}_{P,p^L} =\exp \left( f_L +f_{P,p^L} +\frac{1}{2} [f_L, f_{P,p^L}]+\ldots \right) \nonumber \\
&=& \exp \Bigl( -i(X\cdot P) \otimes (a\cdot P) +i(a\cdot X)P^\alpha \otimes P_\alpha -\frac{i}{2}(X\cdot P)\otimes (a\cdot P)^2 +\frac{i}{2} a^2(X\cdot P)\otimes P^2 +i(a\cdot X)(a\cdot P)P_\alpha \otimes P^\alpha \nonumber \\
&& -\frac{i}{2}(a\cdot X)P^2 \otimes (a\cdot P)+\frac{i}{2}(a\cdot X)P_\alpha \otimes (a\cdot P)P^\alpha -\frac{i}{3}(X\cdot P)\otimes (a\cdot P)^3 +\frac{i}{2}a^2 (X\cdot P)\otimes (a\cdot P)P^2 \nonumber \\
&& +\frac{i}{3} (a\cdot X)P^\alpha \otimes (a\cdot P)^2P_\alpha -\frac{i}{2}(a\cdot X)(a\cdot P)P^2\otimes (a\cdot P) -\frac{i}{2}(a\cdot X)P^2 \otimes (a\cdot P)^2 +i(a\cdot X)(a\cdot P)P^\alpha \otimes (a\cdot P)P_\alpha \nonumber \\
&& +i(a\cdot X)(a\cdot P)^2P_\alpha \otimes P^\alpha -\frac{i}{2}a^2(a\cdot X)P^2P^\alpha \otimes P_\alpha -\frac{i}{2}a^2(a\cdot X)P^\alpha \otimes P^2P_\alpha -\frac{i}{2}(a\cdot X)P_\alpha P_\beta \otimes P^\alpha P^\beta \nonumber \\
&& +O(a_0^4) \Bigr).
\end{eqnarray}
The obtained result coincides with the twist calculated in section \ref{3}.
\subsection{Left covariant and left noncovariant}
In this example, we use the twist for the realization corresponding to the left noncovariant (left ordering of $\kappa$-Minkowski spacetime) and similarity transformations between the two undeformed Heisenberg algebras to reproduce the twist for the left covariant realization. In this case, the terms we obtain using our perturbative method can be summed and we are able to reproduce the analytical result.

The realization corresponding to the left ordering of $\kappa$-Minkowski spacetime is given by $\hat{x}_0= x_0^l$, $\hat{x}_i =x_i^l \exp (-a\cdot p^l)$, where $p^l_\mu$ are the corresponding momenta to $x^l$, ($[p^l_\mu, x^l_\nu]=-i\eta_{\mu\nu}$). The relations between $\{x_\mu^L,p_\mu^L\}$ and $\{x_\mu^l, p_\mu^l\}$ are given by
\begin{equation}
x_i^L=x_i^l, \quad x_0^L=x_0^l e^{a_0p_0^l},\quad x_0^l= x_0^L (1-a_0p_0^L), \quad p_i^L=p_i^l, \quad p_0^L=\frac{1- e^{-a_0p_0^l}}{a_0}, \quad p_0^l=-\frac{1}{a_0}\ln (1-a_0p_0^L). \label{Ll}
\end{equation}
We define the coproduct $\Delta_0'$ such that
\begin{equation}
\Delta_0' p_\mu^l \equiv p_\mu^l \otimes 1+ 1\otimes p_\mu^l, \quad \Delta_0' x_\mu^l \equiv x_\mu^l\otimes 1,
\end{equation}
and, as before, calculate $\Delta_0'$ for $x_\mu^L$ and $p_\mu^L$ using eqs. in \eqref{Ll} and the homomorphism property of the coproduct
\begin{eqnarray}
& \Delta_0' p_i^L &=\Delta_0' p_i^l =p_i^L \otimes 1+ 1\otimes p_i^L =\Delta_0 p_i^L \\
& \Delta_0' x_i^L &=\Delta_0' x_i^l = x_i^L \otimes 1 =\Delta_0 x_i^L \\
& \Delta_0' p_0^L &=\Delta_0' \left( \frac{1- e^{-a_0p_0^l}}{a_0} \right) =p_0^L\otimes 1+1\otimes p_0^L -a_0p_0^L \otimes p_0^L +O(a_0^4) \\
& \Delta_0' x_0^L &=(x_0^L(1-a_0p_0^L) \otimes 1)(\Delta_0'(e^{-a_0p_0^l})) \nonumber \\
& &= x_0^L\otimes 1 +a_0x_0^L\otimes p_0^L +a_0^2x_0^L\otimes (p_0^L)^2 -a_0^3x_0^L \otimes (p_0^L)^3 +O(a_0^4).
\end{eqnarray}
Now, for the relative twist $\mathcal{F}_{L,l}$, defined with
\begin{equation}
\Delta_0' p_\mu^L=\mathcal{F}_{L,l} \Delta_0 p_\mu^L \mathcal{F}_{L,l}^{-1}, \quad \Delta_0' x_\mu^L =\mathcal{F}_{L,l} \Delta_0 x_\mu^L \mathcal{F}_{L,l}^{-1},
\end{equation}
using the method of section \ref{2}, we find the first terms in the expansion $\mathcal{F}_{L,l}=\exp(f_1^{L,l} +f_2^{L,l} +f_3^{L,l} +...)$, $f_k^{L,l} \propto a_0^k$ and then summarize using the induction
\begin{equation}
\mathcal{F}_{L,l} =\exp \left( -ix_0^Lp_0^L\otimes \left( a_0p_0^L+\frac{(a_0p_0^L)^2}{2} +\frac{(a_0p_0^L)^3}{3} +\ldots \right) \right) =\exp (-ix_0^Lp_0^L \otimes \ln Z),
\end{equation}
where the shift operator in the left covariant realization is given by $Z^{-1}=1-a_0p_0^L$. This twist satisfies the cocycle condition. The twist for the realization corresponding to the left ordering of the $\kappa$-Minkowski spacetime is known to be \cite{Meljanac-4, ms06, Govindarajan-1}
\begin{equation}
\mathcal{F}_l= \exp (ix^l_i p^l_i \otimes a_0p_0^l) =\exp (ix^l_i p^l_i \otimes \ln Z) =\exp (ix^L_i p^L_i \otimes \ln Z),
\end{equation}
where the second equality comes from the fact that in the realization corresponding to the left ordering of $\kappa$-Minkowski spacetime $Z=e^{a_0p_0^l}$, and the third one follows from eqs. in \eqref{Ll}. Now we get for the left covariant twist
\begin{equation}
\mathcal{F}_L= \mathcal{F}_{l}\mathcal{F}_{L,l} =\exp \Bigl( i(x_i^L p_i^L -x_0^L p_0^L) \otimes \ln Z\Bigr) =\exp (ix^L\cdot p^L \otimes \ln Z ),
\end{equation}
which is the result known from \cite{Borowiec-3}. Note that $\mathcal{F}_L$ and $\mathcal{F}_l$ are both Drinfeld twists (they satisfy the cocycle condition). Remark: We could have also reproduced both of the twists, $\mathcal{F}_L$ and $\mathcal{F}_l$, using our method from section II. 
Note that  $\mathcal{F}_{P,p^L} = \Delta_0'\mathcal{S} \left(\mathcal{S}^{-1} \otimes \mathcal{S}^{-1}\right)$.

\section{Outlook and Discussion}
In the following we will discuss some of the physical motivations for studying the mathematical structure of $\kappa$-Minkowski space time and its realization via quantum phase space.

$\kappa$-Minkowski spacetime and $\kappa$-Poincar\'{e} algebra may provide a setting for trapping the signals of quantum gravity effects, which may be found in observation of ultrahigh energy cosmic rays, contradicting the usual understanding of electron-positron production in collisions of  high energy photons and other high energy astrophysical processes alike. It turns out that deviations of this kind can be explained by the modified dispersion relations, whose modification can be traced back to  the deformation of spacetime, particularly of the $\kappa$ type \cite{dispersion, dispersion1, bgmp10}.

The main question is what are the main effects of Planck scale physics, that is, how the nature of NC spacetime affects the construction of QFT's. The obvious one is the change in  the particle statistics. The information about particle statistics is encoded in the $R$-matrix. For example, given a free scalar field $\phi$ and knowing $R$-matrix one can modify the algebra of creation and annihilation operators via
\begin{equation}
\phi(x)\otimes\phi(y)-R\phi(y)\otimes\phi(x)=0
\end{equation}
and deform the usual spin-statistics relation of usual bosons at Planck scale.
The $R$ matrix is defined via 
\begin{equation}\begin{split}\label{r}
R=\tilde{\mathcal{F}}\mathcal{F}^{-1}&=1\otimes 1+ia_{0}x^{\alpha}\left(p_{\beta}\otimes\left[\frac{\partial \varphi^{\ \beta}_{\alpha}}{\partial a_{0}}\right]_{a_{0}=0} -\left[\frac{\partial \varphi^{\ \beta}_{\alpha}}{\partial a_{0}}\right]_{a_{0}=0}\otimes p_{\beta}\right)+O(a^{2}_{0})\\
&=1\otimes 1+ia_{0}r+O(a^{2}_{0}),
\end{split}\end{equation}
where $r$ is the classical $r$-matrix. Note that the measurement of the system which involves only single-particle observables will not reveal whether  the system is truly deformed or not. The twisted deformation that we are dealing with here only makes itself manifest in the multi-particle sector.

In \cite{super} a large class of supersymmetric classical $r$-matrices, describing supertwist deformation of Poincare and Euclidian superalgebra was presented. It is interesting to study the extended version (like in \cite{diff})of quantum phase space, that is super-phase space, and its deformation.

Although we have been analyzing only $\kappa$-deformations of Minkowski spacetime, one can write the most general expression for the  twist operator $\mathcal{F}=e^{f_{1}+...}\ $ up to the first order for a general NC space by writing (symbolically) $f_{1}$ as  $f_{1}\equiv\sum G_{i}\otimes G_{j}$, where generators $G_{i}\in\left\{p_{i},p_{0},M_{ij},M_{i0},x_{0}p_{0},x_{k}p_{k}\right\}$ ($M_{\mu\nu}$ are Lorentz generators). Then using (\ref{t}) one can obtain the most general deformations (up to the first order) of the coalgebra of Poincar\'{e}  algebra (namely calculate $\Delta M$ and $\Delta p$). Also we can calculate all possible classical $r$-matrices (similarly as done in eq. (\ref{r})). This analysis is in some sense alternative  to the one carried out in \cite{balesteros}, where (2+1) (A)dS and Poincar\'{e} $r$-matrices are not only defined by $\kappa$-deformations, but also correspond to multiparametric $r$-matrices. It is very interesting, from a physical point of view, to analyze these multiparametric deformations, since in the particular case of (2+1) quantum gravity, it was stated in \cite{komentar} that the perturbations of the vacuum state of a Chern-Simons quantum gravity theory with cosmological constant $\Lambda$ are invariant under transformations that close under a certain deformation of the (A)dS algebra. The low energy regime/zero-curvature limit of this algebra was found to be $\kappa$-Poincar\'{e} algebra \cite{Lukierski-1, Majid-Ruegg}. These multiparametric analysis will provide all possible deformations of (A)dS and Poincar\'{e} algebras .

Recently, \cite{fluid}, using realization formalism of NC spaces (namely for Snyder space), NC fluid was analyzed. The NC fluid generalizes the fluid model in the action functional formulation of the NC space. Fluid equations of motion and their perturbative solutions were derived \cite{fluid1}. It is of interest to further  investigate this line of research using realization formalism for more general NC spaces.

In this paper we have analyzed the structure of both quantum phase space $\mathcal{H}$ and $\kappa$-deformed phase space $\hat{\mathcal{H}}$. Quantum phase space is described with Heisenberg algebra and has Hopf algebroid structure. $\kappa$-deformed phase space has also Hopf algebroid structure.  The coordinates of $\kappa$-Minkowski space are realized in terms of quantum phase space with twisted Hopf algebroid structure. Realizations, i.e. different bases of quantum phase space are related by similarity transformations. We have presented a general method for constructing the twist operator. Using this method, for the first time, we give the twist for natural realization (classical basis), prove the cocycle condition and discuss the corresponding $\kappa$-Poincar\'{e} algebra and $R$-matrix. We believe that the result presented in this paper  are  crucial and necessary mathematical background for analyzing physical theories on NC spaces.  The idea is to construct quantum field theory (especially gauge theory) and gravity in the Hopf algebroid setting generalizing the ideas presented by the group of Wess et al. $R$-matrix will enable us to define particle statistics and to properly quantize fields, while the twist operator will provide the star product, which is crucial for writing the action and deriving the equations of motion.

\appendix
\section{Undeformed Hopf algebroid}
 Undeformed Hopf algebroid is defined by total algebra $\mathcal{H}$ (quantum phase space ), base algebra $\mathcal{A}$, multiplication $m$, coproduct $\Delta_{0}$, antipode $S_{0}$, counit $\epsilon_{0}$, source map $\alpha_{0}$ and target map $\beta_{0}$. The coproduct $\Delta_{0}$ is a mapping $\Delta_{0}: \mathcal{H}\mapsto \mathcal{U}(\mathcal{R}_{0})(\mathcal{A}\otimes\mathcal{A})\Delta_{0}\mathcal{T}/\mathcal{I}_{0}$ defined by (\ref{primitive}). The coproduct $\Delta_{0}$ is a homomorphism, satisfies (\ref{primitive}) and the coassociativity condition
\begin{equation}\label{a1}
(\Delta_{0}\otimes1)\Delta_{0}=(1\otimes\Delta_{0})\Delta_{0}.
\end{equation}
The antipode $S_{0}$ is a maping $S_{0}: \mathcal{H}\mapsto \mathcal{H}$ and an antihomomorphism $S_{0}(h_{1}h_{2})=S_{0}(h_{2})S_{0}(h_{1})$  $\forall$ $h_{1},h_{2}\in\mathcal{H}$. For generators of Heisenberg algebra we have
\begin{equation} \label{a2}
S_{0}(x_{\mu})=x_{\mu}, \quad S_{0}(p_{\mu})=-p_{\mu}.
\end{equation}
 The counit $\epsilon_{0}: \mathcal{H}\mapsto\mathcal{A}$ is defined by $\epsilon_{0}(h)=h\triangleright 1$ $\in\mathcal{A}\subset\mathcal{H}$, $\forall$ $h\in\mathcal{H}$. Note that $\epsilon_{0}(\mathcal{H})=\mathcal{A}$. 
The target map $\alpha_{0}: \mathcal{A}\mapsto\mathcal{H}$ and source map $\beta_{0}: \mathcal{A}\mapsto\mathcal{H}$ are equal and coincide with $\mathcal{A}\hookrightarrow\mathcal{H}$.  The coproduct $\Delta_{0}$, antipode $S_{0}$ and counit $\epsilon_{0}$ satisfy the following relations
\begin{equation}\begin{split} \label{a3}
&m(\epsilon_{0}\otimes1)\Delta_{0}=m(1\otimes\epsilon_{0})\Delta_{0}=1\\
&m(S_{0}\otimes1)\Delta_{0}=m(1\otimes S_{0})\Delta_{0}=\epsilon_{0}.
\end{split}\end{equation}
The coproduct $\Delta_{0}$, antipode $S_{0}$ and counit $\epsilon_{0}$ are highly related  via (\ref{a1}-\ref{a3}). For example, if we consider momentum representation:
\begin{equation}
\epsilon^{\text{mom}}_{0}(p_{\mu})=p_{\mu}, \quad \epsilon^{\text{mom}}_{0}(x_{\mu})=0,
\end{equation}
then 
\begin{equation}
\Delta^{\text{mom}}_{0}(p_{\mu})=p_{\mu}\otimes1, \quad \Delta^{\text{mom}}_{0}(x_{\mu})=x_{\mu}\otimes1+1\otimes x_{\mu},
\end{equation}
and 
\begin{equation}
S^{\text{mom}}_{0}(p_{\mu})=p_{\mu}, \quad S^{\text{mom}}_{0}(x_{\mu})=-x_{\mu}. 
\end{equation}

\section{Twisted Hopf algebroid}
Twisted Hopf algebroid is defined by total algebra $\mathcal{H}$ (quantum phase space ), base algebra $\hat{\mathcal{A}}$ (where for elements of $\hat{\mathcal{A}}$ a particular realization is explicitly used), multiplication $m$, twisted coproduct $\Delta_{\mathcal{F}}\equiv\Delta$, antipode $S_{\mathcal{F}}\equiv S$, counit $\epsilon_{\mathcal{F}}\equiv\hat{\epsilon}$, source map $\hat{\alpha}$ and target map $\hat{\beta}$. This twisted structure also satisfies the axioms of Hopf algebroid. Using twist $\mathcal{F}\in (\mathcal{H}\otimes\mathcal{H})/\mathcal{J}$ and its inverse $\mathcal{F}^{-1}\in (\mathcal{H}\otimes\mathcal{H})/\mathcal{J}_{0}$, that satisfies cocycle condition (\ref{CO}) and normalization condition (\ref{NO}), we define twisted coproduct $\Delta : \Delta_{0}\mathcal{H}\mapsto\Delta\mathcal{H}$
\begin{equation}
\Delta h=\mathcal{F}\Delta_{0}h\mathcal{F}^{-1}, \quad \forall \ \ h\in\mathcal{H}
\end{equation}
which satisfies the coassociativity condition
\begin{equation}\label{codel}
(\Delta\otimes1)\Delta=(1\otimes\Delta)\Delta .
\end{equation}
The antipode $S: \mathcal{H}\mapsto\mathcal{H}$ is an antihomomorphism defined by
\begin{equation}
S(h)=\chi S_{0}(h)\chi^{-1}
\end{equation}
where $\chi^{-1}=m\left[(S_{0}\otimes1)\mathcal{F}^{-1}\right]$. The counit $\hat{\epsilon}:\mathcal{H}\mapsto\hat{\mathcal{A}}\subset{\mathcal{H}}$ is defined by
\begin{equation}
\epsilon(h)=m\left\{\mathcal{F}^{-1}(\triangleright\otimes1)(\epsilon_{0}(h)\otimes1)\right\}.
\end{equation}
So, we have
\begin{equation}\begin{split}
 \hat{\epsilon}(f)=\hat{f},& \quad \epsilon_{0}(\hat{f})=f,\\
\hat{\epsilon}(f\star g)=\hat{f}\ \hat{g},& \quad \epsilon_{0}(\hat{f}\ \hat{g})=f\star g,\\
 \hat{\epsilon}(\epsilon_{0}(\hat{f}))=\hat{f},& \quad \epsilon_{0}(\hat{\epsilon}(f))=f,  
\end{split}\end{equation} 
$\forall f\equiv f(x),\ g\equiv g(x)\in\mathcal{A}$ and $\hat{f}\equiv\hat{f}(\hat{x}),\ \hat{g}\equiv\hat{g}(\hat{x})\in\hat{\mathcal{A}}$. Let us reconstruct the source and the target maps from the twist.
First, we define \an\ and \bn, $\an:\As\rightarrow\Ah\subset\Hn$,
$\bn:\As\rightarrow\Hn$ by
\begin{equation}
  \an(f(x))=m\left(\mathcal{F}^{-1}(\rhd\otimes 1)(\ano(f(x))\otimes1)\right),\;\;
  \ano(f(x))=f(x)
\end{equation}
and
\begin{equation}
  \bn(f(x))=m\left(\tilde{\mathcal{F}}^{-1}(\rhd\otimes 1)
  (\bno(f(x))\otimes1)\right),\;\;\bno(f(x))=f(x).
\end{equation}
Now, the source and the target maps are given by
\begin{equation}
  \ah=\an\eno|_{\Ah}
\end{equation}
and
\begin{equation}
  \bh=\bn\eno|_{\Ah}.
\end{equation}  The coproduct $\Delta$, antipode $S$ and counit $\epsilon$ satisfy the following relations
\begin{equation}\begin{split}
&m(\hat{\epsilon}\otimes1)\Delta=m(1\otimes S^{-1}\hat{\epsilon}\ \ S)\Delta=1\\
&\ \ \ \ \ \ \ \ m(S\otimes1)\Delta=S^{-1}\hat{\epsilon}\ \ S\\
&\ \ \ \ \ \ \ \ \ \ \  m(1\otimes S)\Delta=\hat{\epsilon}
\end{split}\end{equation}
which are compatible with the Hopf algebroid structure in \cite{Lu}. The relation between Lu's paper \cite{Lu} and our approach is $\alpha\epsilon\mapsto\hat{\epsilon}$ and $\beta\epsilon\mapsto S^{-1}\hat{\epsilon}$, since in our case $\hat{\epsilon}:\mathcal{H}\mapsto\hat{\mathcal{A}}\subset\mathcal{H}$. In the undeformed case the relation is $\alpha\epsilon\mapsto\epsilon_{0}$ and $\beta\epsilon\mapsto S^{-1}_{0}\epsilon_{0}=\epsilon_{0}$.

\section{Hopf algebroid structure of $\hat{\mathcal{H}}$}
$\kappa$-deformed phase space $\hat{\mathcal{H}}$ generated by NC coordinates $\hat{x}_{\mu}$ and momentum $p_{\mu}$ also has Hopf algebroid structure which is defined by the total algebra $\hat{\mathcal{H}}$, base algebra $\hat{\mathcal{A}}\subset\hat{\mathcal{H}}$,  multiplication map $m$, coproduct $\Delta$ (satisfying (\ref{codel}) and (\ref{delxkapa})), antipode $S$, counit $\hat{\epsilon}$, source map $\hat{\alpha}$ and target map $\hat{\beta}$ (see \cite{mali} for the construction related to bicrossproduct basis). The counit $\hat{\epsilon}$ is defined by
\begin{equation}
\hat{\epsilon}(\hat{h})=\hat{h}\blacktriangleright 1 , \quad \forall\hat{h}\in\hat{\mathcal{H}}.
\end{equation}

Let us introduce $\hat{y}_{\mu}$ as the right multiplication by $\hat{x}_{\mu}$
\begin{equation}
\hat{y}_{\mu}\blacktriangleright\hat{f}(\hat{x})=\hat{f}(\hat{x})\hat{x}_{\mu}
\end{equation}
and 
\begin{equation}
\Delta \hat{y}_{\mu}=1\otimes\hat{y}_{\mu}
\end{equation} with the property 
\begin{equation}
[\hat{x}_{\mu},\hat{y}_{\nu}]=0, \quad [\hat{y}_{\mu},\hat{y}_{\nu}]=-i(a_{\mu}\hat{y}_{\nu}-a_{\nu}\hat{y}_{\mu}).
\end{equation}
Note that relation $\hat{Q}_{\mu}=\hat{y}_{\mu}\otimes 1 -1\otimes\hat{x}_{\mu}$ with the property $\hat{Q}_{\mu}\blacktriangleright\hat{\mathcal{A}}\otimes\hat{\mathcal{A}}=0$ generates the right ideal $\mathcal{J}$ defined by
 $\mathcal{J}=\mathcal{U}_{+}(\hat{Q})\hat{\mathcal{H}}\otimes\hat{\mathcal{H}}$, which satisfies $\mathcal{J}\blacktriangleright\hat{\mathcal{A}}\otimes\hat{\mathcal{A}}=0$.\\
The antipode $S$ is defined by $S(\hat{y}_{\mu})=\hat{x}_{\mu}$ and satisfies
\begin{equation}\begin{split}
&m(\hat{\epsilon}\otimes1)\Delta=m(1\otimes S^{-1}\hat{\epsilon}\ \ S)\Delta=1\\
&\ \ \ \ \ \ \ \ m(S\otimes1)\Delta=S^{-1}\hat{\epsilon}\ \ S\\
&\ \ \ \ \ \ \ \ \ \ \  m(1\otimes S)\Delta=\hat{\epsilon}
\end{split}\end{equation}
The antipode for $p_{\mu}$, $S(p_{\mu})$ follows from  $m(S\otimes1)\Delta(p_{\mu})=m(1\otimes S)\Delta(p_{\mu})=0$. The source map $\hat{\alpha}: \hat{\mathcal{A}}\mapsto\hat{\mathcal{H}}$ is a homomorphism and the target map $\hat{\beta}: \hat{\mathcal{A}}\mapsto\hat{\mathcal{H}}$ is an antihomomorphism defined by $\hat{\beta}=S^{-1}\hat{\alpha}$.
Coproduct $\Delta$, antipode $S$, counit $\hat{\epsilon}$ and multiplication map $m$ provide the Hopf algebroid structure of $\hat{\mathcal{H}}$ that is isomorphic to $\kappa$-deformed twisted Hopf algebroid structure of $\mathcal{H}$ presented in Appendix B.

Note that twist $\mathcal{F}$ obtained from the realization (\ref{xrealization}) leads to 
\begin{equation}
\hat{x}_{\mu}=m\left(\mathcal{F}^{-1}(\triangleright\otimes1)( x_{\mu}\otimes 1)\right)=x_{\alpha}\varphi^{\alpha}_{\ \mu}(p).
\end{equation}
We can also define $\hat{y}_{\mu}$
\begin{equation}
\hat{y}_{\mu}=m\left(\tilde{\mathcal{F}}^{-1}(\triangleright\otimes1)(x_{\mu}\otimes1)\right)=x_{\alpha}\tilde{\varphi}^{\alpha}_{\ \mu}(p),
\end{equation}
with $\tilde{\mathcal{F}}$ defined by $\tilde{\mathcal{F}}=\tau_0\mathcal{F}\tau_0$, where $\tau_0$ is the flip operator, $\tau_0 (h_1\otimes h_2)= h_2 \otimes h_1, \forall h_1, h_2 \in \mathcal{H}$.

\noindent{\bf Acknowledgment}\\
We would like to thank  K. S. Gupta, D. Kova\v{c}evi\'{c},  A. Pachol, A. Samsarov and Z. \v{S}koda   for useful comments and discussions. S.M. would like to thank A. Borowiec for discussions and  J. Lukierski for useful comments.
This work was supported by the Ministry of Science and Technology of the Republic
of Croatia under contract No. 098-0000000-2865. R.\v{S}. gratefully acknowledges support from the DFG within the Research Training Group 1620 ``Models of Gravity''.\\

\end{document}